\documentclass[prc,showpacs,nofootinbib,preprint]{revtex4-1}
\usepackage{epsfig,graphics,color} 
\usepackage{subfigure} 
\usepackage{bbm} 
\usepackage{mathtools} 
\usepackage{amssymb} 
\usepackage{mathrsfs} 
\usepackage{amsmath} 
\usepackage[T1]{fontenc}
\usepackage{palatino}   
\DeclareMathAlphabet{\mathpzc}{OT1}{pzc}{m}{it}
\usepackage{txfonts}  
\usepackage[pdftex]{hyperref}
\hypersetup{%
  pdftitle={Large-Nc sum rules}
  ,pdfauthor={Yonggoo Heo}
  ,pdfsubject={Axial couplings up to Q3}
  ,colorlinks
}
\usepackage{pdfsync}
\usepackage{tcolorbox,soul}
\usepackage{hhline}
\usepackage{lineno}
\usepackage{feynmp}
\allowdisplaybreaks[1]
\definecolor{annr}{rgb}{0.86, 0.08, 0.24} 
\definecolor{anng}{rgb}{0.13, 0.55, 0.14}  
\definecolor{annb}{rgb}{0.0, 0.28, 0.67}   
\definecolor{mono}{rgb}{0.65, 0.65, 0.65} 

\usepackage{pifont}

\usepackage{array}
\newcolumntype{L}{>{$}l<{$}}
\newcolumntype{C}{>{$}c<{$}}
\newcolumntype{R}{>{$}r<{$}}
\newcommand{\Slash}[1]{\ooalign{\hfil/\hfil\crcr$#1$}}   
\newcommand{\der}{\partial}

\newcommand{\dd}{\mathrm{d}}
\newcommand{\tr}{\textbf{tr}}

\newtcbox{\colorBox}[1]{
  nobeforeafter,colframe=#1,colback=#1!10!white,
  boxrule=0.5pt,arc=4pt,boxsep=0pt,
  left=3pt,right=3pt,top=0pt,bottom=0pt,
  tcbox raise base
}

\begin{document}
\title{The chiral Lagrangian with three flavors and large-$N_c$ sum rules
}
\author{Yonggoo Heo$^1$, C. Kobdaj$^1$\footnote{
    \hypertarget{correspondingAuthors}{\text{Corresponding authors.}}
} and Matthias F.M. Lutz$^{2 \,\hyperlink{correspondingAuthors}{\rm a}}$}
\affiliation{$^1$ Suranaree University of Technology, Nakhon Ratchasima, 30000, Thailand}
\affiliation{$^2$ GSI Helmholtzzentrum f\"ur Schwerionenforschung GmbH,\\
  Planck Str. 1, 64291 Darmstadt, Germany}
\date{\today}
\begin{abstract}
We reconsider the chiral Lagrangian with three-flavor baryon fields.
A systematic analysis of all LEC that contribute to the axial-vector and pseudoscalar currents in the baryon octet and decuplet fields at next-to-leading order is performed. While there are 4 LEC relevant at leading order, the number 
of relevant LEC at subleading chiral order is 23. For those a leading order large-$N_c$ analysis predicts 3 and 18 sum rules respectively. At the next accuracy level the number of sum rules is reduced to 2 and 8. Our results are illustrated by a tree-level analysis of available axial-vector coupling constants and strong decay widths of the baryon decuplet states. 
\end{abstract}

\maketitle
\tableofcontents

\newpage
\section{Introduction}

The chiral Lagrangian with three light flavors is strongly bivalent to its potential and challenges \cite{Weinberg:1979poi,Gasser:1984gg,Krause:1990xc,Bernard:1993nj,Kaiser:1996js,Fearing:1999fw,Lutz:2001yb,Kolomeitsev:2003kt,Oller:2007qd,CaroRamon:1999jf,Fernando:2017yqd,Fernando:2019upo,Heo:2019cqo,Holmberg:2019ltw,Sauerwein:2021jxb}. 
While the flavor structure of its interaction terms at a given chiral order is highly predictive, its slow convergence properties ask for the quantitative control of higher order terms. The number of terms relevant at a given order gets quickly prohibitive large. 

In the baryon sector of QCD, an approach invented by 't Hooft and Witten 
~\cite{Hooft:1973jz,Witten:1979kh,Luty:1993fu,Dashen:1994qi} offers a remedy by considering the number of colors ($N_c$) in QCD 
as a free parameter, in terms of its inverse a controllable expansion parameter is available. 
A specific approach how to derive large-$N_c$ sum rules for the set of low-energy constants (LEC) in the 
chiral Lagrangian systematically has been established more recently in~\cite{Lutz:2010se,Heo:2019cqo}. 
Those sum rules provide a huge parameter reduction and therefore are instrumental in studies of 
chiral dynamics with three light flavors in QCD. 

For instance, a significant chiral extrapolation of the baryon octet and decuplet masses became 
possible at N$^3$LO  ~\cite{Lutz:2010se,Lutz:2014oxa,Lutz:2018cqo} and was successfully confronted with the available lattice data set ~\cite{Aubin:2004wf,WalkerLoud:2008bp,Aoki:2008sm,Lin:2008pr,Durr:2008zz,Alexandrou:2009qu,Durr:2011mp,WalkerLoud:2011ab}. The analysis relies heavily on the various large-$N_c$ sum rules for the LEC. Recently, this approach has been systematically extended to further classes of LEC relevant at chiral order $Q^3$  ~\cite{Heo:2019cqo}.  The 24 symmetry-conserving terms were analyzed in the $1/N_c$ expansion and sets of $22$ and $19$ sum rules valid at leading and subleading orders respectively were established. 

The current study is to complete our study of the chiral SU(3) Lagrangian at $Q^3$. We will consider a complete list of symmetry-breaking terms in the Lagrangian at that order.  
Here we apply the framework that has been established in~\cite{Luty:1993fu,Dashen:1994qi,Lutz:2010se,Heo:2019cqo}. We find sum rules amongst such $23$ LEC.
In addition, we illustrate their role at hand of the empirical values for the $\beta$ decays of the baryon octet states and the $\pi$ decays of the decuplet states ~\cite{Dashen:1994qi,Dai:1995zg,FloresMendieta:1998ii}. We note that in such a study  it is necessary to discriminate axial-vector from pseudoscalar currents. This was not the target of the previous works \cite{Dai:1995zg,FloresMendieta:1998ii} and needs clarification.

The work is organized as follows.
In Section~\ref{sec:chi-lagrangian-AS},
we review the chiral SU(3) Lagrangian at order $Q^3$ with the symmetry-breaking fields involving the quark masses of QCD. Matrix elements of one- and two-body quark currents as relevant in our study are presented in Section~\ref{sec:correlation-function:LHS}. Then in the main part, Section~\ref{sec:correlation-function:RHS}, 
sum rules for the LEC are derived according to the large-$N_c$-operator analysis. 
We illustrate the role of the sum rules in terms of a phenomenological tree-level study in Section~\ref{sec:numerics}.
The paper is closed with a summary.

\newpage 
\section{Chiral Lagrangian with baryon octet and decuplet fields}
\label{sec:chi-lagrangian-AS}
The low-energy constants (LEC) at leading order (LO) to study interactions involving octet and decuplet baryons in the chiral perturbation theory are introduced with the following Lagrangian,
\begin{eqnarray}
  \label{Lag:Q1}
  {\mathscr L}^{(1)}
  & = &
  \tr\,\bar{B}\,(i\Slash{D} - M_{[8]})\,{B}
  \nonumber\\ &&
  -\, \tr\,(\bar{B}_\mu\cdot[
  (i\Slash{D} - M_{[10]})\,g^{\mu\nu} - i(\gamma^\mu\,D^\nu +\gamma^\nu\,D^\mu)
  + \gamma^\mu\,(i\Slash{D} + M_{[10]})\,\gamma^\nu
  ]\,{B}_\nu)
  \nonumber\\ &&
  +\, F\,\tr\,\bar{B}\,\gamma^\mu\gamma_5\,[{iU}_\mu,\,{B}]_-
  + D\,\tr\,\bar{B}\,\gamma^\mu\gamma_5\,[{iU}_\mu,\,{B}]_+
  \nonumber\\ &&
  +\, C\,\tr\,\{(\bar{B}^\mu\cdot {iU}_\mu)\,B -\bar{B}\,({iU}_\mu^\dagger\cdot{B}^\mu) \}
  - H\,\tr\,(\bar{B}^\mu\cdot\gamma^\nu\gamma_5\,{B}_\mu)\,{iU}_\nu
  \,,
\end{eqnarray}
where $[A,\,B]_\pm = A\,B \pm B\,A$.
The kinetic terms for both baryons are presented in~(\ref{Lag:Q1}) as well, in order to make it complete at this chiral order.
$M_{[8]}$ and $M_{[10]}$ denote the baryon masses in the chiral SU(3) limit.
The convenient \textit{dot-notation} suggested in~\cite{Lutz:2001yb} for baryon decuplet fields is recalled as
\begin{align}
  \label{def:dot-notation}
  (\bar{B}^{\bar\tau} \cdot {B}_\tau)^i_j
  = \bar{B}^{\bar\tau}_{jkl}\,{B}_\tau^{ikl}
  \,,\quad
  (\bar{B}^{\bar\tau} \cdot \varphi)^i_j = \epsilon^{kli}\,\bar{B}^{\bar\tau}_{knj}\,\varphi^n_l
  \,,\quad
  (\varphi \cdot {B}_\tau)^i_j = \epsilon_{klj}\,\varphi^l_n\,{B}_\tau^{kni}
  \,,
\end{align}
where the $\varphi$ stands for any octet and the flavor indices $i$, $j$, $k$, $l$, $m$, $n=1,\,2,\,3$.
The chiral building block ${U}_\lambda$ in which the \textit{Nambu-Goldstone} (NG) boson field $\Phi$ is  encoded as well as the classical source functions, $l_\lambda\equiv v_\lambda - a_\lambda$ and $r_\lambda\equiv v_\lambda + a_\lambda$ of QCD~\cite{Gasser:1984gg},
and the covariant derivatives of baryon states are given as
\begin{eqnarray}
  {U}_\lambda
  & = &
  \tfrac{1}{2}\,{u}^\dagger\,\left(
    \der_\lambda{e}^{i\Phi/f}
    + {e}^{i\Phi/f}\,il_\lambda
    - ir_\lambda{e}^{i\Phi/f}
  \right)\,{u}^\dagger
  \,,\qquad
  {u}=e^{i\Phi/2f}
  \,,\nonumber\\
  (D_\lambda {B})^{i}_{j}
  & = &
  \der_\lambda {B}^{i}_{j} + \Gamma^{i}_{\lambda,k}\,{B}^{k}_{j} + {B}^{i}_{k}\,\Gamma^{\dagger k}_{\lambda,j}
  \,,\nonumber\\
  (D_\lambda {B}_\tau)^{ijk}
  & = &
  \der_\lambda {B}_\tau^{ijk}
  + \Gamma^{i}_{\lambda,l}\,{B}_\tau^{ljk}
  + \Gamma^{j}_{\lambda,l}\,{B}_\tau^{ilk}
  + \Gamma^{k}_{\lambda,l}\,{B}_\tau^{ijl}
  \,,\nonumber\\
  \Gamma_\lambda
  & = &
  \tfrac{1}{2}\,{u}\,(\der_\lambda - il_\lambda)\,{u}^\dagger
  + \tfrac{1}{2}\,{u}^\dagger\,(\der_\lambda - ir_\lambda)\,{u}
  \,,
\end{eqnarray}
where the chiral connection $\Gamma_\lambda^\dagger=-\Gamma_\lambda$.

We recall the conventions for the chiral Lagrangian as used in the current work~\cite{Krause:1990xc,Lutz:2010se,Lutz:2001yb,Lutz:2018cqo},
in which the NG boson $\Phi$ and the baryon octet $B$ fields are decomposed into isospin multiplets as follows.
\begin{eqnarray}
  \label{def:fields-isospin-decomp}
  \Phi
  & = &
  \Phi^a\,\lambda_a
  =
  \vec\tau\cdot\vec\pi + \alpha^\dag\cdot{K}
  + {K}^\dag \cdot \alpha + \eta\,\lambda_8
  \,,\nonumber\\
  \sqrt{2}\,B
  & = &
  B^a\,\lambda_a
  =
  \alpha^\dag\cdot{N} + \lambda_8\Lambda
  + \vec\tau\cdot\vec\Sigma + {\Xi}^T i\sigma^2 \cdot\alpha
  \,,\nonumber\\
  && \,
  \sqrt{2}\,\alpha^\dag
  =
  (\lambda_4 + i\lambda_5 ,\,\lambda_6 + i\lambda_7)
  \,,\qquad
  \vec\tau = (\lambda_1,\,\lambda_1,\,\lambda_3)
  \,,
\end{eqnarray}
where the matrices $\lambda_a$ are the standard Gell-Mann generators of the SU(3) algebra
and the $\sigma^2$ is one of the Pauli matrices.
The baryon decuplet field $B_{\tau}^{ijk}$ comes with three completely symmetric flavor indices $i$, $j$, $k$ as
\begin{flalign}
  \label{def:field-representation-decuplet}
  &&&
  B_{\tau}^{111} = \Delta^{++}_\tau
  \,,&&
  B_{\tau}^{112} = \Delta^{+}_\tau/\sqrt3
  \,,&&
  B_{\tau}^{122} = \Delta^{0}_\tau/\sqrt3
  \,,&&
  B_{\tau}^{222} = \Delta^{-}_\tau
  \,,& \nonumber\\ &&&
  B_{\tau}^{113} = \Sigma^{*+}_\tau/\sqrt3
  \,,&&
  B_{\tau}^{123} = \Sigma^{*0}_\tau/\sqrt6
  \,,&&
  B_{\tau}^{223} = \Sigma^{*-}_\tau/\sqrt3
  \,,&&& \nonumber\\ &&&
  B_{\tau}^{133} = \Xi^{*0}_\tau/\sqrt3
  \,,&&
  B_{\tau}^{233} = \Xi^{*-}_\tau/\sqrt3
  \,,&& &&& \nonumber\\ &&&
  B_{\tau}^{333} = \Omega^{-}_\tau
  \,,&& && &&&
\end{flalign}
where the components are identified with the states in the particle basis for convenience.

Derivation of correlations amongst the low-energy parameters of the chiral Lagrangian as they follow from a $1/N_c$ expansion is the main goal of this work. For that purpose, we consider axial, scalar, and pseudoscalar currents from QCD,
\begin{eqnarray}
  \label{eq:QCD:VASP}
  A^\mu_a(x)
  & = &
  \tfrac{1}{2}\,\bar\Psi(x)\,\gamma^\mu\gamma_5\,\lambda_a\,\Psi(x)
  \,,\quad
  S_b(x) = \tfrac{1}{2}\,\bar\Psi(x)\,\lambda_b\,\Psi(x)
  \,,\quad
  P_b(x) = \tfrac{1}{2}\,\bar\Psi(x)\,i\gamma_5\,\lambda_b\,\Psi(x)
  \,,
\end{eqnarray}
where the $\Psi(x)$ represents the Heisenberg quark-field operators and the Gell-mann flavor matrices $\lambda_a$ with the supplementary singlet matrix $\lambda_0=\sqrt{2/3}\,{\bf 1}_{\!3\times 3}$ are recalled.
Given the chiral Lagrangian, it is well defined how to derive the contribution to such matrix elements in baryon states.~\cite{Gasser:1983yg,Gasser:1984gg,Ecker:1989yg}. 
The matrices of classical source fields of QCD, axial $a_\mu$, scalar $s$, and pseudoscalar $p$, come in with the building blocks
\begin{eqnarray}
  \label{eq:vasp}
  i\,U_\mu
  & = &
  a_\mu
  - \tfrac{1}{2\,f}\,\der_\mu\Phi
  + \cdots
  \,,\nonumber\\
  \chi_\pm
  & = &
  \tfrac{1}{2}\,(u\,\chi_0^\dagger\,u \pm u^\dagger\chi_0\,u^\dagger)
  =
  \Bigg\{\begin{array}{rcl}
    \chi_+
    & = &
    + 2\,B_0\,s
    + \chi_0
    + \cdots
    \,,\\
    \chi_-
    & = &
    - 2\,B_0\,ip
    + \tfrac{i}{2\,f}\,\lbrack \Phi,\,\chi_0\rbrack_+
    + \cdots
    \,,
  \end{array}
\end{eqnarray}
where we put $B_0=1/2$ from now on for notational simplicity.
Details will be studied in the following sections, however,
it would be worth bringing in the sum rules of LEC from the previous work~\cite{Lutz:2018cqo,Dashen:1993jt,Dashen:1994qi}.
Based on the chiral Lagrangian~(\ref{Lag:Q1}), the large-$N_c$ sum rules are
\begin{eqnarray}
  \label{res-FDCHs:LO}
  &&
  F= \tfrac{2}{3}\,D \,,\qquad C=2\,D \,,\qquad H = 3\,D\,,
  \qquad
  \text{at LO}
  \,,\\
  \label{res-FDCHs:subLO}
  &&
  C=2\,D\,,\qquad H= 9\,F-3\,D \,,
  \qquad\qquad \quad \,\,\,\,
  \text{at sub-LO}
  \,.
\end{eqnarray}
Those parameters, $F$, $D$, $C$, and $H$, play an important role in the chiral theory,
such as the meson-baryon coupled-channel interactions~\cite{Lutz:2001yb}
and the chiral extrapolation for baryon ground-state masses~\cite{Lutz:2018cqo}.
Some of their values can be readily identified with $\beta$ decays of octet states~\cite{Okun_2014,Dai:1995zg,FloresMendieta:1998ii}
and with $\pi$ decays of decuplet states~\cite{Dai:1995zg,FloresMendieta:1998ii,Bernard:2003xf} at this level of chiral order.

There are contributions to those parameters at higher order arising, when the following chiral Lagrangian is taken into account.
\begin{eqnarray}
  \label{eq:cal:Lag:Q3}
  {\mathscr L}^{(3)}_{\chi}
  & = &
  {\mathscr L}^{(3)}_{\chi_+} + {\mathscr L}^{(3)}_{\chi_-}
  \,,\\
  \label{eq:cal:Lag:Q3:chi+}
  {\mathscr L}^{(3)}_{\chi_+}
  & = &
  \tfrac{1}{2}\,{F}_{0}\,\big\{
  \tr\,\bar{B}\,\gamma_\mu\gamma_5\,[iU^\mu,\,[\chi_+,\,{B}]_+]_+
  + {\rm h.c.}
  \big\}
  + \tfrac{1}{2}\,{F}_{1}\,\big\{
  \tr\,\bar{B}\,\gamma_\mu\gamma_5\,[iU^\mu,\,[\chi_+,\,{B}]_-]_+
  + {\rm h.c.}
  \big\}
  \nonumber\\ && \,
  +\, \tfrac{1}{2}\,{F}_{2}\,\big\{
  \tr\,\bar{B}\,\gamma_\mu\gamma_5\,[iU^\mu,\,[\chi_+,\,{B}]_+]_-
  + {\rm h.c.}
  \big\}
  + \tfrac{1}{2}\,{F}_{3}\,\big\{
  \tr\,\bar{B}\,\gamma_\mu\gamma_5\,[iU^\mu,\,[\chi_+,\,{B}]_-]_-
  + {\rm h.c.}
  \big\}
  \nonumber\\ && \,
  +\, \tfrac{1}{2}\,F_4\,\big\{
  \tr\,\bar{B}\,\gamma_\mu\gamma_5\,{B}\,\tr\,\chi_+\,iU^\mu
  + {\rm h.c.}
  \big\}
  + \tfrac{1}{2}\,F_5\,\big\{
  \tr\,\bar{B}\,\gamma_\mu\gamma_5\,iU^\mu\,\tr\,\chi_+\,{B}
  + {\rm h.c.}
  \big\}
  \nonumber\\ && \,
  +\, F_6\,\tr\,\bar{B}\,\gamma_\mu\gamma_5\,[iU^\mu,\,{B}]_-\,\tr\,\chi_+
  \nonumber\\ && \,
  +\, {C}_{0}\,\big\{
  \tr\,(\bar{B}_\mu\cdot\,[\chi_+,\,iU^\mu]_+)\,{B}
  + {\rm h.c.}
  \big\}
  + {C}_{1}\,\big\{
  \tr\,(\bar{B}_\mu\cdot\,[\chi_+,\,iU^\mu]_-)\,{B}
  + {\rm h.c.}
  \big\}
  \nonumber\\ && \,
  +\, {C}_{2}\,\big\{
  \tr\,(\bar{B}_\mu\cdot iU^\mu)\,[\chi_+,\,{B}]_+
  + {\rm h.c.}
  \big\}
  + {C}_{3}\,\big\{
  \tr\,(\bar{B}_\mu\cdot iU^\mu)\,[\chi_+,\,{B}]_-
  + {\rm h.c.}
  \big\}
  \nonumber\\ && \,
  +\, {C}_{4}\,\big\{
  \tr\,(\bar{B}_\mu\cdot\chi_+)\,[iU^\mu,\,{B}]_-
  + {\rm h.c.}
  \big\}
  \nonumber\\ && \,
  -\, {H}_{0}\,\tr\,(\bar{B}_\mu\cdot\gamma_\nu\gamma_5\,{B}^\mu)\,[\chi_+,\,iU^\nu]_+
  - {H}_{1}\,\tr\,(\bar{B}_\mu\cdot\gamma_\nu\gamma_5\,{B}^\mu)\,[\chi_+,\,iU^\nu]_-
  \nonumber\\ && \,
  -\, \tfrac{1}{2}\,{H}_{2}\,\big\{
  \tr\,(\bar{B}_\mu\cdot\chi_+)\,\gamma_\nu\gamma_5\,(iU^\nu\cdot{B}^\mu)
  + {\rm h.c.}
  \big\}
  - \tfrac{1}{2}\,{H}_{3}\,\big\{
  \tr\,(\bar{B}_\mu\cdot\gamma_\nu\gamma_5\,{B}^\mu)\,\tr\,\chi_+\,iU^\nu
  + {\rm h.c.}
  \big\}
  \nonumber\\ && \,
  -\, {H}_{4}\,\tr\,(\bar{B}_\mu\cdot\gamma_\nu\gamma_5\,{B}^\mu)\,iU^\nu\,\tr\,\chi_+
  \,,\\
  \label{eq:cal:Lag:Q3:chi-}
  {\mathscr L}^{(3)}_{\chi_-}
  & = &
  {M}_{[8]}\,F_7\,\tr\,\bar{B}\,\gamma_5\,\lbrack\chi_-,\,{B}\rbrack_-
  + {M}_{[8]}\,F_8\,\tr\,\bar{B}\,\gamma_5\,\lbrack\chi_-,\,{B}\rbrack_+
  \nonumber\\ &&\,
  +\, C_5\,\big\{
  \tr\,(\bar{B}^\mu\cdot\chi_-)\,{iD}_\mu{B}
  + {\rm h.c.}
  \big\}
  + {M}_{[8]}\,F_9\,\tr\,\bar{B}\,\gamma_5\,{B}\,\tr\,\chi_-
  \nonumber\\ &&\,
  -\, {M}_{[10]}\,H_5\,\tr\,(\bar{B}_\tau\cdot\gamma_5\,{B}^\tau)\,\chi_-
  - {M}_{[10]}\,H_6\,\tr\,(\bar{B}_\tau\cdot\gamma_5\,{B}^\tau)\,\tr\,\chi_-
  \,,
\end{eqnarray}
whose decuplet sectors are now completed as their octets~\cite{Lutz:2001yb,Oller:2007qd,Frink:2006hx,Holmberg:2018dtv,Goity:1999by} and
$M_{[8]}$ and $M_{[10]}$ denote the baryon masses in the chiral SU(3) limit as in (\ref{Lag:Q1}).
We emphasize that our list here is free of redundant terms.
This was not the case in the previous work~\cite{Lutz:2001yb} where an overcomplete list was constructed.

\vskip-0.15ex 
We have  $7+5+5+6=23$ independent parameters at this chiral order.
Taking the chiral Lagrangian~(\ref{eq:cal:Lag:Q3:chi+})-(\ref{eq:cal:Lag:Q3:chi-}) into account lies on the same line of the previous study~\cite{Heo:2019cqo},
where the other low-energy parameters at the same chiral order as in~(\ref{eq:cal:Lag:Q3:chi+})-(\ref{eq:cal:Lag:Q3:chi-}) were investigated.
Now, we have all independent terms at $Q^3$ having both breaking and preserving the chiral symmetry,
which are related to meson-baryon scattering processes at this chiral order.
The reader should be reminded that the parameters, $F_i$, $C_i$, and $H_i$, contribute to matrix elements of the QCD's axial and pseudoscalar currents $A^\mu_a$ and $P_a $ with Eq.~(\ref{eq:vasp}) alongside contributions of $F$, $D$, $C$, and $H$ from~(\ref{Lag:Q1}).
Further quark currents of QCD, which probe the symmetry-breaking terms directly, will be a part of our current study. As a consequence we will arrive at large-$N_c$ sum rules for the LEC $F_i$, $C_i$, and $H_i$.

\section{Correlation functions from the chiral Lagrangian}
\label{sec:correlation-function:LHS}
The Fourier transformed QCD's currents for the time-ordered axial current alone, product of it with scalar current, and pseudoscalar current itself can be written as
\begin{eqnarray}
  \label{def:VASP}
{\mathcal  A}^{\mu}_{a}(q)
  & = &
  i\int\dd^4x\,e^{iq\cdot x}\,{\mathcal T}\,A^\mu_a(x)
  \,, \qquad \qquad
  \mathcal{S}^{\mu}_{ab}(q)
  = 
  i\int\dd^4x\,e^{iq\cdot x}\,{\mathcal T}\,A^\mu_a(x)\,S_b(0)
  \,,  
  \nonumber\\
  {\mathcal P}_{b}(q)
  & = &
  i\int\dd^4x\,e^{iq\cdot x}\,{\mathcal T}\,P_b(x)
  \,,
\end{eqnarray}
where $A^\mu_a(x)$, $S_b(x)$, and  $P_b(x)$ are defined in Eq.~(\ref{eq:QCD:VASP}).
The baryon matrix elements can be derived from the chiral Lagrangian by using the classical source functions in the building blocks, $U_\mu$ and $\chi_\pm$, in conjunction with their expansions in power of the meson field $\Phi$, as shown in Eq.~(\ref{eq:vasp}).
For our purposes it suffices to evaluate the matrix elements in the strict flavor SU(3) limit with baryon octet and decuplet states,
\begin{eqnarray}
  \label{eq:baryon-states:chiral}
  &&
  |p,\chi,c\rangle
  \,,\qquad \qquad
  |p,\chi,klm\rangle
  \,.
\end{eqnarray}
Each is specified by its three-momentum $p$ and the flavor indices $c=1,\cdots,8$ or $k$, $l$, $m=1,\,2,\, 3$. The spin polarization label is $\chi=1,2$ for the octet or $\chi=1,\cdots,4$ for the decuplet states.
As studied in our earlier work~\cite{Lutz:2010se,Heo:2019cqo}, we focus on the space components of such currents.
Once it is completed,
the large-$N_c$ sum rules for the low-energy constants can be determined.

Let us start with the matrix elements for the pseudoscalar current from the chiral Lagrangian densities~(\ref{Lag:Q1}) and (\ref{eq:cal:Lag:Q3}). The rational behind this strategy is the simplicity of such a correlation function. Only a small subset of LEC contribute here. 
In the nonrelativistic and flavor limit~\cite{Lutz:2010se,Heo:2019cqo} we present them explicitly as
\begin{eqnarray}
  \label{summary:<08|P|08>-LHS-c=0-L3--nre}
  \langle\bar{p},\bar\chi,{d}\,|\,{\mathcal P}_{b}(\bar{p}-{p})\,|\,{p},\chi,c\rangle
  & = &
  - Q^i\,\sigma^i_{\bar\chi\chi}\,\big\{
  \delta_{b0}\,\big(
  \sqrt{3/2}\,F_9
  + \sqrt{2/3}\,F_8
  \big)\,\delta_{{d}{c}}
  \nonumber\\ && \,
  + (1-\delta_{b0})\,\big(
  F_8\,d_{b{c}{d}}
  + F_7\,if_{b{c}{d}}
  \big)
  \big\}
  + \cdots
  \,,\\
  \label{summary:<10|P|10>-LHS-c=0-L3--nre}
  \langle\bar{p},\bar\chi,nop\,|\,{\mathcal P}_{b}(\bar{p}-{p})\,|\,{p},\chi,klm\rangle
  & = &
  - Q^i\,(S^j\,\sigma^i\,S^{j\dagger})_{\bar\chi\chi}\,
  \tfrac{1}{2}\,\delta^{nop}_{xyz}\,\big\{
  \delta_{b0}\,\big(
  \sqrt{6}\,H_6
  + \sqrt{2/3}\,H_5
  \big)\,\delta^{xyz}_{klm}
  \nonumber\\ && \,
  + (1-\delta_{b0})\,H_5\,\Lambda^{b,xyz}_{klm}
  \big\} + \cdots
  \,,\\
  \label{summary:<10|P|08>-LHS-c=0-L3--nre}
  \langle\bar{p},\bar\chi,nop\,|\,{\mathcal P}_{b}(\bar{p}-{p})\,|\,{p},\chi,c\rangle
  & = &
  Q^j\,S^j_{\bar\chi\chi}\,(1-\delta_{b0})\,\tfrac{1}{\sqrt{2}}\,
  C_5\,\Lambda^{nop}_{bc}
  + \cdots
  \,,
\end{eqnarray}
where $Q^i=(\bar{p}-{p})^i/2$.   We emphasize that only terms from (\ref{eq:cal:Lag:Q3:chi-})
turn relevant here. In Eqs.~(\ref{summary:<08|P|08>-LHS-c=0-L3--nre})-(\ref{summary:<10|P|08>-LHS-c=0-L3--nre}),
the summation over the indices occurring twice in a single term is understood with $i$, $j$, $x$, $y$, $z=1,\,2,\,3$.
The dots represent additional terms that are suppressed as the 3-momenta $\bar{p}$ and ${p}$ approach zero.
The $\sigma^i$ are the Pauli matrices for spin $1/2$ and the $S^i$ are the spin-transition matrices from $1/2$ to $3/2$.
Those spin structures emerge from the Dirac bilinears in the nonrelativistic limit~\cite{Lutz:2010se,Heo:2019cqo}.
We recall some useful flavor structures 
\begin{eqnarray}
\label{def:flavor-transition-matrices}
  && \delta^{{n}{o}{p}}_{klm}
   = 
  \tfrac{1}{6}\,(
  \delta_{{n}k}\,\delta_{{o}l}\,\delta_{{p}m}
  + \delta_{{n}l}\,\delta_{{o}m}\,\delta_{{p}k}
  + \delta_{{n}m}\,\delta_{{o}k}\,\delta_{{p}l}
  + \delta_{{n}l}\,\delta_{{o}k}\,\delta_{{p}m}
  + \delta_{{n}m}\,\delta_{{o}l}\,\delta_{{p}k}
  + \delta_{{n}k}\,\delta_{{o}m}\,\delta_{{p}l}
  )
  \,,\nonumber\\
&&  \Lambda^{a,nop}_{klm}
   = 
  (\lambda_a)_{nr}\,\delta^{rop}_{klm}
  \,,\qquad
  \Lambda^{ab}_{klm}
  =
  \epsilon_{ijn}\,(\lambda_a)_{io}\,(\lambda_b)_{jp}\,\delta^{nop}_{klm}
  \,,\qquad
  \Lambda^{{n}{o}{p}}_{ab}
  =
  \delta^{{n}{o}{p}}_{klm}\,\epsilon_{kij}\,(\lambda_a)_{li}\,(\lambda_b)_{mj}
  \,,
\end{eqnarray}
where the other flavor indices run correspondingly, i.e.
$a$, $b=1,\cdots,8$ and
$i$, $j$, $k$, $l$, $m$, $n$, $o$, $p=1,\,2,\,3$. 
The absence of the singlet in the  off-diagonal sector is obviously from $(\bar{B}^{\bar\tau} \cdot {\bf 1}_{3\times 3})^i_j = \epsilon^{kli}\,\bar{B}^{\bar\tau}_{knj}\,\delta^n_l=0$ from Eq.~(\ref{def:dot-notation}).


We continue with the matrix elements of the time-ordered product of the product of axial-vector and scalar currents as~(\ref{def:VASP}). Based on the chiral Lagrangian~(\ref{Lag:Q1}) and (\ref{eq:cal:Lag:Q3}), we find in the nonrelativistic and flavor limit~\cite{Lutz:2010se,Heo:2019cqo} the result
\begin{eqnarray}
  \label{summary:<08|AS|08>-LHS-c=0-L3+-nre}
  \langle\bar{p},\bar\chi,{d}\,|\,{\mathcal S}^i_{ab}(\bar{p}-p)\,|\,p,\chi,c\rangle
  & = &
  - \tfrac{1}{8}\,\sigma^i_{\bar\chi\chi}\,\big\{
  \delta_{b0}\,\sqrt{\tfrac{2}{3}}\,\big[
  8\,{F}_{0}\,d_{ac{d}}
  + \big(
  8\,{F}_{2}
  + 12\,F_6
  \big)\,if_{ac{d}}
  \big]
  \nonumber\\ && \,
  +\, (1-\delta_{b0})\,\big[
  \big(
  4\,{F}_{0}
  - 4\,{F}_{3}
  + 2\,F_5
  \big)\,\big(
  \delta_{ac}\,\delta_{b{d}} + \delta_{a{d}}\,\delta_{bc}
  \big)
  \nonumber\\ && \quad
  +\, \big(
  - 4\,{F}_{1}
  + 4\,{F}_{2}
  \big)\,\big(
  if_{ace}\,d_{b{d}e}
  - if_{a{d}e}\,d_{bce}
  \big)
  \nonumber\\ && \quad
  +\, \big(
  \tfrac{4}{3}\,{F}_{0}
  + 4\,{F}_{3}
  + 4\,F_4
  \big)\,\delta_{ab}\,\delta_{c{d}}
  + \big(
  - 4\,{F}_{0}
  + 12\,{F}_{3}
  \big)\,d_{abe}\,d_{c{d}e}
  \nonumber\\ && \quad
  +\, 8\,{F}_{1}\,d_{abe}\,if_{c{d}e}
  \big]
  \big\}
  + \cdots
  \,,\\
  \label{summary:<10|AS|10>-LHS-c=0-L3+-nre}
  \langle\bar{p},\bar\chi,nop\,|\,{\mathcal S}^i_{ab}(\bar{p}-p)\,|\,p,\chi,klm\rangle
  & = &
  - \tfrac{1}{4}\,(S^j\sigma^iS^{j\dagger})_{\bar\chi\chi}\,\delta^{nop}_{rst}\,\big\{
  \delta_{b0}\,\sqrt{\tfrac{2}{3}}\,\big(
  2\,{H}_{0}
  + 3\,{H}_{4}
  \big)\,\Lambda^{a,rst}_{klm}
  \nonumber\\ && \,
  +\, (1-\delta_{b0})\,\big[
  \big(
  \tfrac{4}{3}\,{H}_{0}
  + {H}_{2}
  + 2\,{H}_{3}
  \big)\,\delta_{ab}\,\delta^{rst}_{klm}
  \nonumber\\ && \quad
  +\, \big(
  2\,{H}_{0}
  + \tfrac{3}{2}\,{H}_{2}
  \big)\,d_{abe}\,\Lambda^{e,rst}_{klm}
  - 2\,{H}_{1}\,if_{abe}\,\Lambda^{e,rst}_{klm}
  \nonumber\\ && \quad
  -\, \tfrac{3}{4}\,{H}_{2}\,\big(
  \Lambda^{a,rst}_{xyz}\,\Lambda^{b,xyz}_{klm} + \Lambda^{b,rst}_{xyz}\,\Lambda^{a,xyz}_{klm}
  \big)
  \big]
  \big\}
  + \cdots
  \,,\\
  \label{summary:<10|AS|08>-LHS-c=0-L3+-nre}
  \langle\bar{p},\bar\chi,nop\,|\,{\mathcal S}^i_{ab}(\bar{p}-p)\,|\,p,\chi,c\rangle
  & = &
  - \tfrac{1}{4\,\sqrt{2}}\,S^i_{\bar\chi\chi}\,\big\{
  \delta_{b0}\,\sqrt{\tfrac{2}{3}}\,\big(
  2\,{C}_{0}
  + 2\,{C}_{2}
  \big)\,\Lambda^{nop}_{ac}
  \nonumber\\ && \,
  +\, (1-\delta_{b0})\,\big[
  \big(
  - 2\,{C}_{0}
  + {C}_{2}
  - {C}_{3}
  - {C}_{4}
  \big)\,d_{abe}\,\Lambda^{nop}_{ce}
  \nonumber\\ && \quad
  +\, \big(
  2\,{C}_{1}
  + {C}_{2}
  - {C}_{3}
  + {C}_{4}
  \big)\,if_{abe}\,\Lambda^{nop}_{ce}
  \nonumber\\ && \quad
  +\, \big(
  {C}_{2}
  - {C}_{3}
  + 2\,{C}_{4}
  \big)\,(d_{ace} + i\,f_{ace})\,\Lambda^{nop}_{be}
  \nonumber\\ && \quad
  +\, \big(
  2\,{C}_{3}
  - {C}_{4}
  \big)\,(d_{bce}+if_{bce})\,\Lambda^{nop}_{ae}
  \big]
  \big\}
  + \cdots
  \,,
\end{eqnarray}
which illustrates the relevance of this correlation function for a specific subset of LEC.
The repeated indices run accordingly as $e=1,\cdots,8$ and
$j$, $x$, $y$, $z$, $r$, $s$, $t=1,\,2,\,3$. 
Again the leading order 
LEC $F$, $D$, $C$, and $H$ do not contribute. Only terms from (\ref{eq:cal:Lag:Q3:chi+}) contribute.

We turn to  the matrix elements for the axial current from the chiral Lagrangian densities~(\ref{Lag:Q1}) and~(\ref{eq:cal:Lag:Q3}).
This time we relax the condition of the flavor limit and assume the isospin limit with $m_u= m_d$ only.
This is useful for simple applications of our expressions further below in this section. 
In the nonrelativistic limit~\cite{Lutz:2010se,Heo:2019cqo} we present them explicitly as
\begin{eqnarray}
  \label{summary:<08|A|08>-LHS-c=0-L3+-nre}
  \langle\bar{p},\bar\chi,{d}\,|\,{\mathcal  A}^i_a(0)\,|\,p,\chi,c\rangle
  & = &
  \sigma^i_{\bar\chi\chi}\,\big\{
  F\,if_{ac{d}}
  + D\,d_{ac{d}}
  + \tfrac{1}{3}\,\varepsilon_0\,\big[
  2\,F_0\,d_{ac{d}}
  + \big(
  2\,F_2
  + 3\,F_6
  \big)\,if_{ac{d}}
  \big]
  \nonumber\\ && \,
  -\, \tfrac{1}{\sqrt{3}}\,\varepsilon_8\,\big[
  \big(
  \tfrac{1}{3}\,F_0
  + F_3
  + F_4
  \big)\,\delta_{8a}\,\delta_{c{d}}
  + \big(
  - F_0
  + 3\,F_3
  \big)\,d_{8ae}\,d_{ec{d}}
  \nonumber\\ && \quad
  +\, F_1\,if_{8ce}\,d_{e{d}a}
  - F_1\,if_{8{d}e}\,d_{eca}
  + F_2\,d_{8ce}\,if_{e{d}a}
  - F_2\,d_{8{d}e}\,if_{eca}
  \nonumber\\ && \quad
  +\, \big(
  F_0
  - F_3
  + \tfrac{1}{2}\,F_5
  \big)\,\big(
  \delta_{8c}\,\delta_{a{d}}
  + \delta_{8{d}}\,\delta_{ac}
  \big)
  \big]
  \big\}
  + \cdots
  \,,\\
  \label{summary:<10|A|10>-LHS-c=0-L3+-nre}
  \langle\bar{p},\bar\chi,nop\,|\,{\mathcal  A}^i_a(0)\,|\,p,\chi,klm\rangle
  & = &
  (S^j\,\sigma^i\,S^{j\dagger})_{\bar\chi\chi}\,\delta^{nop}_{xyz}\,\big\{
  \tfrac{1}{2}\,H\,\Lambda^{a,xyz}_{klm}
  + \tfrac{1}{6}\,\varepsilon_0\,\big(
  2\,H_{0}
  + 3\,H_{4}
  \big)\,\Lambda^{a,xyz}_{klm}
  \nonumber\\ && \,
  -\, \tfrac{1}{2\,\sqrt{3}}\,\varepsilon_8\,\big[
  \big(
  \tfrac{4}{3}\,H_{0}
  + \tfrac{5}{6}\,H_{2}
  + 2\,H_{3}
  \big)\,\delta_{8a}\,\delta^{xyz}_{klm}
  \nonumber\\ && \quad
  +\, \big(
  2\,H_{0}
  + 2\,H_{2}
  \big)\,d_{8ae}\,\Lambda^{e,xyz}_{klm}
  + 2\,H_{1}\,if_{8ae}\,\Lambda^{e,xyz}_{klm}
  \nonumber\\ && \quad
  -\, \tfrac{3}{8}\,H_{2}\,\big(
  \Lambda^{a,xyz}_{rst}\,\Lambda^{8,rst}_{klm} + \Lambda^{8,xyz}_{rst}\,\Lambda^{a,rst}_{klm}
  \big)
  \big]
  \big\}
  + \cdots
  \,,\\
  \label{summary:<10|A|08>-LHS-c=0-L3+-nre}
  \langle\bar{p},\bar\chi,nop\,|\,{\mathcal A}^i_a(0)\,|\,p,\chi,c\rangle
  & = &
  {S}^i_{\bar\chi\chi}\,\big\{
  \tfrac{1}{2\sqrt{2}}\,C\,\Lambda^{nop}_{ac}
  + \tfrac{1}{6\sqrt{2}}\,\varepsilon_0\,\big(
  2\,C_0
  + 2\,C_2
  )\,\delta^{nop}_{xyz}\,\Lambda^{xyz}_{ac}
  \nonumber\\ && \,
  -\, \tfrac{1}{2\sqrt{6}}\,\varepsilon_8\,\big[
  \big(
  2\,C_0\,d_{8ae}
  + 2\,C_1\,if_{8ae}
  \big)\,\delta^{nop}_{xyz}\,\Lambda^{xyz}_{ec}
  + 2\,C_4\,if_{ace}\,\delta^{nop}_{xyz}\,\Lambda^{xyz}_{8e}
  \nonumber\\ && \quad
  +\, \big(
  2\,C_2\,d_{8ce}
  + 2\,C_3\,if_{8ce}
  \big)\,\delta^{nop}_{xyz}\,\Lambda^{xyz}_{ae}
  \big]
  \big\}
  + \cdots
  \,,
\end{eqnarray}
with the help of other chiral parameters
\begin{eqnarray}
  \label{eq:other-chiral-parameters}
    \chi_0
  & = &
  \tfrac{1}{3}\,\varepsilon_0\,{\bf 1}_{3\times 3}
  - \tfrac{1}{\sqrt{3}}\,\varepsilon_8\,\lambda_8 \,,  \qquad \qquad \quad \quad m = m_u = m_d \,,
  \nonumber\\
  \varepsilon_0
  & = & 2\,B_0\,(2\,m +m_s )\simeq 
  2\,m_K^2+m_\pi^2
  \,,\qquad
  \varepsilon_8
  =2\,B_0\,(m_s-m) \simeq 
  2\,(m_K^2-m_\pi^2)
  \,,
\end{eqnarray}
which decomposes $\chi_0$ into two parts in accordance with the \textit{Gell-Mann-Oakes-Renner} relation~\cite{GellMann:1968rz}.

There are some certain points that we can directly obtain from this result.
First of all,
the customary axial couplings $g_A$ for $\beta$ decays of octet baryons can be directly obtained by using the matrix elements of QCD's current ${\mathcal  A}^i_a$~(\ref{summary:<08|A|08>-LHS-c=0-L3+-nre}).
A specific process is described by the linear combination of two axial currents ${\mathcal  A}^i_a$ with indices $i$ and $a$ selected correspondingly.
Our results are consistent with Eqs.~(119) and~(121) of~\cite{Lutz:2001yb}.
The possible minus sign in $\Xi^-$ channel originated in the different convention of ours for the isospin decomposition~(\ref{def:fields-isospin-decomp}) from others is compensated in Tab.~\ref{tab:gA:beta:math}.
Each $g_A$ can be then written as a linear combination of $F$, $D$, and $F_i$. Streamlined expressions are obtained in terms of $F_R$ and $D_R$ with
\begin{eqnarray}
\label{def-FR-DR}
  F_R
  & = &
  F
  + \varepsilon_0\,\big(F_6 + \tfrac{2}{3}\,F_2\big)
  \,,\qquad \qquad \qquad 
  D_R
  =
  D
  + \varepsilon_0\,\big(\tfrac{2}{3}\,F_0\big) \,.
\end{eqnarray}

\begin{table}[t]
  \centering
\begin{tabular}{L@{$\,\rightarrow\,$}L@{$\,:\,$}RRRL}
  {n}
  &{p}\,{e}^-\bar\nu_e
  &g_A=
  &F_R&+D_R
  &
  +\varepsilon_8\,\big[
  +\tfrac{1}{3}\,F_{0}-F_1+\tfrac{1}{3}\,F_2-F_3
  \big]
  \,,\!\!\!\nonumber\\
  \Sigma^-
  &\Lambda\,{e}^-\bar\nu_e
  &g_A=
  &&\sqrt{\tfrac{2}{3}}\,D_R
  &
  +\tfrac{1}{\sqrt{6}}\,\varepsilon_8\,\big[
  -\tfrac{4}{3}\,F_{0}-F_5
  \big]
  \,,\!\!\!\nonumber\\ \hline
  \Lambda
  &{p}\,{e}^-\bar\nu_e
  &g_A=
  &-\sqrt{\tfrac{3}{2}}\,F_R&-\sqrt{\tfrac{1}{6}}\,D_R
  &
  -\tfrac{1}{\sqrt{6}}\,\varepsilon_8\,\big[
  \tfrac{11}{6}\,F_{0}-\tfrac{1}{2}\,F_1+\tfrac{3}{2}\,F_2-\tfrac{3}{2}\,F_3+F_5
  \big]
  \,,\!\!\!\nonumber\\
  \Sigma^-
  &{n}\,{e}^-\bar\nu_e
  &g_A=
  &-F_R&+D_R
  &
  +\varepsilon_8\,\big[
  -\tfrac{1}{6}\,F_{0}-\tfrac{1}{2}\,F_1+\tfrac{1}{6}\,F_2+\tfrac{1}{2}\,F_3
  \big]
  \,,\!\!\!\nonumber\\
  \Xi^-
  &\Lambda\,{e}^-\bar\nu_e
  &g_A=
  &
  \sqrt{\tfrac{3}{2}}\,F_R&-\sqrt{\tfrac{1}{6}}\,D_R
  &
  -\tfrac{1}{\sqrt{6}}\,\varepsilon_8\,\big[
  \tfrac{11}{6}\,F_{0}+\tfrac{1}{2}\,F_1-\tfrac{3}{2}\,F_2-\tfrac{3}{2}\,F_3+F_5
  \big]
  \,,\!\!\!\nonumber\\
  \Xi^0
  &\Sigma^+\,{e}^-\bar\nu_e
  & g_A=
  &
  F_R&+D_R
  &
  +\varepsilon_8\,\big[
  -\tfrac{1}{6}\,F_{0}+\tfrac{1}{2}\,F_1-\tfrac{1}{6}\,F_2+\tfrac{1}{2}\,F_3
  \big]
  \,
\end{tabular}
\caption{
  The axial coupling $g_A$ for $\beta$ decays of octet baryons are decomposed into low-energy parameters,
  where $F_R$ and $D_R$ are introduced in Eq.~(\ref{def-FR-DR}).
  The horizontal separation indicates $g_A$ between $\Delta S=0,1$. 
}
\label{tab:gA:beta:math}
\end{table}

\begin{table}[b]
  \centering
\begin{tabular}{L@{$\,\rightarrow\,$}L@{$\!:\;\;\;$}RRRL}
  \Delta
  &N\,\pi \qquad \phantom{xx}
  & g_P=\,
  & \big\{C_A&
  &
  +\, \varepsilon_8\,\big[
  -\tfrac{2}{3}\,C_{0}+\tfrac{1}{3}\,C_{2}-C_{3}
  \big]
  \big\}
  \,,\!\!\!\nonumber\\
  \Sigma^*
  &\Lambda\,\pi
  &g_P=
  & -\frac{1}{\sqrt{2}}\,\big\{C_A&
  &
  +\, \varepsilon_8\,\big[
  -\tfrac{2}{3}\,C_{0}+\tfrac{2}{3}\,C_{2}
  \big]
  \big\}
  \,,\!\!\!\nonumber\\
  \Sigma^*
  &\Sigma\,\pi
  &g_P=
  &-\frac{1}{\sqrt{3}}\, \big\{C_A&
  &
  +\, \varepsilon_8\,\big[
  -\tfrac{2}{3}\,C_{0}-\tfrac{2}{3}\,C_{2}-2\,C_{4}
  \big]
  \big\}
  \,,\!\!\!\nonumber\\
  \Xi^*
  &\Xi\,\pi
  &g_P=
  & -\frac{1}{\sqrt{2}}\,\big\{C_A&
  &
  +\, \varepsilon_8\,\big[
  -\tfrac{2}{3}\,C_{0}+\tfrac{1}{3}\,C_{2}+C_{3}-C_{4}
  \big]
  \big\}
  \,.
\end{tabular}
\caption{
  The hadronic coupling constants $g_P$ as used in (\ref{def-gP}) are decomposed into low-energy parameters.
}
\label{tab:gP:pion:math}
\end{table}

Given our notations it is useful to provide tree-level expressions for hadronic three-point coupling constants also.
For simplicity we focus on  $g_{\pi{N}{N}}$, $g_{\bar{K}{N}\Lambda}$, and $g_{\bar{K}{N}\Sigma}$ here and relate those to corresponding axial coupling constants.
We can arrive at the following \textit{Dashen-Weinstein} type relations~\cite{Dashen:1970vh}
\begin{eqnarray}
  \label{eq:Dashen-Weinstein-v00}
  \frac{f}{{M}_{N}}\,g_{\pi{N}{N}}
  - g_A^{{n}\rightarrow {p}\,{e}^-\bar\nu_e}
  & = &
  - 2\,m_\pi^2\,\frac{{M}_{[8]}}{2\,{M}_{N}}\,(F_8+F_7)
  \,,\nonumber\\
  \frac{f}{{M}_{N}+{M}_\Lambda}\,\sqrt{2}\,g_{\bar{K}{N}\Lambda}
  - g_A^{\Lambda\rightarrow {p}\,{e}^-\bar\nu_e}
  & = &
  \frac{2\,m_K^2}{\sqrt{6}}\,\frac{{M}_{[8]}}{{M}_{N}+{M}_\Lambda}\,(F_8+3\,F_7)
  \,,\nonumber\\
  \frac{2\,f}{{M}_{N}+{M}_\Sigma}\,g_{\bar{K}{N}\Sigma}
  - g_A^{\Sigma^-\rightarrow {n}\,{e}^-\bar\nu_e}
  & = &
  - 2\,m_K^2\,\frac{{M}_{[8]}}{{M}_{N}+{M}_\Sigma}\,(F_8-F_7)
  \,.
\end{eqnarray}

Further hadronic coupling constants $g_P$ are required in the description of
isospin-averaged $\pi$ decays of a decuplet baryon $B_i$ state. The isospin-averaged decay width is
\begin{eqnarray}
  \label{def-gP}
&&  \Gamma_{{B}_{i}\rightarrow{B}_{f} \pi}
  = 
  \frac{E+{M}_{f}}{2\,\pi\,f^2}\,\frac{p^3}{12\,{M}_{i}}\,
 g_P^2
  \,,\qquad \qquad \quad
  p
  =\frac{\surd([{M}_{i}^{2}-({M}_{f}+{m}_\pi)^{2}]\,[{M}_{i}^{2}-({M}_{f}-{m}_\pi)^{2}])}{2\,{M}_{i}}
  \,,\nonumber\\ &&
  E+\surd({m}_\pi^2+p^2)
  ={M}_{i}=
  \surd({M}_{f}^2+p^2)+\surd({m}_\pi^2+p^2)
  \,,
\end{eqnarray}
where ${M}_{i}$ and ${M}_{f}$ are for masses of considered decuplet and octet states, respectively. 
The various values for $g_P$ are listed in Tab.~\ref{tab:gP:pion:math} in terms of our set of LEC and the 
particularly useful combination
\begin{eqnarray}
 C_A = C + \tfrac{2}{3}\,\varepsilon_0 \,\Big( C_0 +C_2 + C_5 \Big)
 -  \tfrac{2}{3}\,\varepsilon_8 \, C_5 \,.
 \label{def-CA}
\end{eqnarray}
Tab.~\ref{tab:gA:beta:math} and Tab.~\ref{tab:gP:pion:math} will be instrumental in the illustration of our large-$N_c$ sum rules to be established in the following section.

\section{Low-energy constants under large-$N_c$ QCD}
\label{sec:correlation-function:RHS}
Sum rules for low-energy constants can be derived from QCD correlation functions expanded in terms of the large-$N_c$ operators as
\begin{eqnarray}
  \label{def-largeN-expansion}
  \langle\,\bar{p},\bar\chi\,|\,{\mathcal O}_{QCD}\,|\,{p},\,\chi\,\rangle
  =
  \sum_{n=0}^\infty\,c_n(\bar{p},{p})\,
  (\,\bar\chi\,|\,{\mathcal O}^{(n)}_{\rm static}\,|\,\chi\,)
  \,,
\end{eqnarray}
where flavor indices are suppressed for simplicity.
In this expansion, (\ref{def:VASP}) may serve as a specific example for ${\mathcal O}_{QCD}$ at $N_c = 3$ with the physical baryon states, $|\,{p},\,\chi\,\rangle$~(\ref{eq:baryon-states:chiral}), that carry the three-momentum $p$. On the other hand, the effective baryon states, $|\,\chi\,)$, do not. It is important to bring in that all dynamical information in (\ref{def-largeN-expansion}) is moved into appropriate coefficient functions $c_n(\bar{p},{p})$ being free of either the flavor or the spin quantum numbers of the initial or the final baryon state. The contributions on its right-hand-side can be sorted according to their relevance at large values of $N_c$. 
The effective baryon states, $|c,\chi)$ and $|klm,\chi)$, have a mean-field structure that can be generated by effective quark operators which satisfy the bosonic commutation relations in spin-flavor space~\cite{Dashen:1994qi}.
They correspond to the baryon states already introduced with (\ref{eq:baryon-states:chiral}) for the particular choice $N_c =3$. A complete set of color-neutral one-body operators may be constructed in terms of the very same static quark operators~\cite{Lutz:2001yb,Lutz:2010se,Heo:2019cqo}. For this physical case where there is a flavor octet with spin-one-half, $|c,\chi)$,
or a flavor decuplet with spin-three-half, $|klm,\chi)$, we recall the well-established results.
\begin{eqnarray}
  \label{result:one-body-operators}
  {\mathbbm 1}\,|\,c,\chi)
  & = &
  3\,|\,c,\chi)
  \,,\nonumber\\
  {\mathbbm 1}\,|\,klm,\chi)
  & = &
  3\,|\,klm,\chi)
  \,,\nonumber\\
  J^i\,|\,c,\chi)
  & = &
  \tfrac{1}{2}\,\sigma^i_{\bar\chi\chi}\,|\,c,\bar\chi)
  \,,\nonumber\\
  J^i\,|\,klm,\chi)
  & = &
  \tfrac{3}{2}\,(S^j\,\sigma^i\,S^{j\dagger})_{\bar\chi\chi}\,|\,klm,\bar\chi)
  \,,\nonumber\\
  T_a\,|\,c,\chi)
  & = &
  if_{ace}\,|\,e,\chi)
  \,,\nonumber\\
  T_a\,|\,klm,\chi)
  & = &
  \tfrac{3}{2}\,\Lambda^{a,xyz}_{klm}\,|\,xyz,\chi)
  \,,\nonumber\\
  G^i_a\,|\,c,\chi)
  & = &
  \tfrac{1}{2}\,\sigma^i_{\bar\chi\chi}\,(d_{ace} + \tfrac{2}{3}\,if_{ace})\,|\,e,\bar\chi)
  + \tfrac{1}{2\sqrt{2}}\,S^i_{\bar\chi\chi}\,\Lambda^{xyz}_{ac}\,|\,xyz,\bar\chi)
  \,,\nonumber\\
  G^i_a\,|\,klm,\chi)
  & = &
  \tfrac{3}{4}\,(S^j\,\sigma^i\,S^{j\dagger})_{\bar\chi\chi}\,\Lambda^{a,xyz}_{klm}\,|\,xyz,\bar\chi)
  + \tfrac{1}{2\sqrt{2}}\,S^{i\dagger}_{\bar\chi\chi}\,\Lambda^{ae}_{klm}\,|\,e,\bar\chi)
  \,,
\end{eqnarray}
with the orthonormal conditions
\begin{eqnarray}
  \label{eq:orthonormalities}
  ({d},\bar\chi\,|\,c,\chi)
  & = &
  \delta_{\bar\chi\chi}\,\delta_{{d}c}
  \,,\quad
  (nop,\bar\chi\,|\,klm,\chi)
  =
  \delta_{\bar\chi\chi}\,\delta^{nop}_{klm}
  \,,\quad
  (nop,\bar\chi\,|\,c,\chi)
  = 0
  \,.
\end{eqnarray}

There are infinitely many terms one may write down in the large-$N_c$ operator expansion~(\ref{def-largeN-expansion}).
The static operators ${\mathcal O}_{\rm static}^{(n)}$ are finite products of the one-body operators $J^i$, $T_a$, and $G^i_a$. 
In contrast, the counting of $N_c$ factors is intricate,
since there is a subtle balance of suppression and enhancement effects.
An $r$-body operator consisting of the $r$ products of any of the spin and flavor operators receives the suppression factor $N_c^{-r}$~\cite{Luty:1993fu,Dashen:1994qi}.
This is counteracted by enhancement factors for the flavor and spin-flavor operators, $T_a$ and $G^i_a$,
that are produced by taking baryon matrix elements at $N_c \neq 3$.
This leads to the enhancement factors~\cite{Dashen:1993jt,Dashen:1994qi}
\begin{eqnarray}
  \label{effective-counting}
  J^i \sim N_c^0
  \,, \qquad \quad
  T_a \sim N_c
  \,, \qquad \quad
  G^i_a \sim N_c
  \,.
\end{eqnarray}
Along with the suppression factor $N_c$ from the expansion coefficients in Eq.~(\ref{def-largeN-expansion}), this implies the effective scaling $J^i\sim 1/N_c$ and $T_a\sim G^i_a\sim N_c^0$~\cite{Lutz:2010se,Heo:2018mur}.
According to (\ref{effective-counting}) there are an infinite number of terms contributing at a given order in the $1/N_c$ expansion. Taking higher products of flavor and spin-flavor operators does not reduce the $N_c$ scaling power. A systematic $1/N_c$ expansion is made possible by a set of operator identities that allows a systematic summation of the infinite number of relevant terms.
As a consequence of the SU(2$\otimes N_f$) Lie algebra with $N_f=3$, any commutator of one-body operators can be expressed in terms of one-body operators again. Therefore it suffices to consider anticommutators of the one-body operators supplemented  by a set of additional reduction rules~\cite{Dashen:1993jt,Dashen:1994qi,Lutz:2010se,Heo:2019cqo}.

In this work we focus on the correlation functions as introduced in Eq.~(\ref{def:VASP}).
For the one-body axial current, the operator analysis was done and well-studied in the previous work~\cite{Lutz:2018cqo,Dashen:1993jt,Dashen:1994qi} at leading order for $F$, $D$, $C$, and $H$.
Corresponding results were recalled in Eqs.~(\ref{res-FDCHs:LO})-(\ref{res-FDCHs:subLO}). While an extension of such an analysis may be possible for the subleading counter terms $F_i$, $C_i$ and $H_i$, that would require an extension of the operator analysis in the presence of flavor breaking effects. The previous related work~\cite{Dai:1995zg} can be seen in this context. We note, however, that the expansion strategy used by Dai et al, is distinct to our framework and also the role of their sum rules as impact to the chiral Lagrangian is only implicit. This was already illustrated in our previous study \cite{Lutz:2018cqo} in which the symmetry-breaking counter terms of chiral order $Q^4$ were analyzed. We confirm that this is also the case for our current study, for which we derived from \cite{Dai:1995zg} the following sum rules
\begin{eqnarray}
      &&     F_{0} = 3\,F_1-F_2+3\,F_3
      \,,\qquad
      F_5 = -2\,F_1+\tfrac{10}{3}\,F_2-4\,F_3\,, \qquad C_{2} = 2\,C_{0}+3\,C_{3} \,.
      \label{res-Dai}
\end{eqnarray}
We emphasize that Ref.~\cite{Dai:1995zg} does not provide any information on the LEC $F_{7-9}, H_{5,6}$, and $C_5$,
which couple to the pseudoscalar current only.
As mentioned before we do not expect to recover~(\ref{res-Dai}) in our framework also.

\begin{table}[t]
  \centering
  \begin{tabular}{R@{\;=\;}L@{\,\,}|R@{\;=\;}L@{\,\,}}
    F_7
    & -\frac{1}{3}\, \hat{h}_{1}-\hat{h}_{3}
    &
    F_9
    & \frac{1}{3}\, \hat{h}_{1}-\frac{1}{\sqrt{6}}\,\hat{h}_{2}
    \\
    F_8
    & -\frac{1}{2}\, \hat{h}_{1}
    &
    H_6
    & \frac{1}{2}\, \hat{h}_{1}-\sqrt{\frac{3}{2}}\,\hat{h}_{2}+3\,\hat{h}_{3}
    \\
    C_5
    & \frac{1}{2}\, \hat{h}_{1}
    &
    H_5
    & -\frac{3}{2}\, \hat{h}_{1}-9\,\hat{h}_{3}
    \\
  \end{tabular}
  \caption{
    The low-energy constants introduced in~(\ref{eq:cal:Lag:Q3:chi-}) are correlated with the intermediate parameters of~(\ref{eq:P-ansatz}) via the large-$N_c$ operator expansion~(\ref{def-largeN-expansion}).
  }
  \label{tab:ps+vps}
\end{table}

We return to our framework, in which we find it more efficient to use further correlation functions instead of the axial current alone. In this case a flavor symmetric analysis suffices.   
Consider the simplest case involving the pseudoscalar current. In application of the well established construction rules \cite{Dashen:1993jt,Dashen:1994qi,Lutz:2010se} at leading and subleading orders there are 3 operators only:
\begin{eqnarray}
  \label{eq:P-ansatz}
  {\mathcal P}_b
  & = & \tfrac{1}{2} \,(\bar{p}-{p})^l\,
  \Big\{
  \delta_{b0}\,\hat{h}_{2}\,J^l
  + (1-\delta_{b0})\,\big(
   \hat{h}_{1}\,G^l_b + \hat{h}_{3}\,\lbrack J^l,\,T_b\rbrack_+
  \big)
  \Big\}
  \,,
\end{eqnarray}
where the summation index $l$ runs over the 3-momentum, from $1$ to $3$.
Parameters $\hat{h}_{2,3}$ represent subleading orders according to the 
effective large-$N_c$ scaling of ours.
In application of our previous work~\cite{Lutz:2010se} the ansatz in 
Eq.~(\ref{eq:P-ansatz}) and the QCD's pseudoscalar currents~(\ref{summary:<08|P|08>-LHS-c=0-L3--nre})-(\ref{summary:<10|P|08>-LHS-c=0-L3--nre}) are readily matched, with results collected in Tab.~\ref{tab:ps+vps}.
There are $6-1=5$ sum rules at LO with $\hat{h}_{2,3}=0$ in the $1/N_c$ expansion:
\begin{eqnarray}
  \label{eq:sumrule:chi-:LO}
  &&\,
  F_7=\tfrac{2}{3}\,F_8
  \,,\quad
  C_5=-F_8
  \,,\quad
  H_5=3\,F_8
  \,,\quad
  F_9=-\tfrac{2}{3}\,F_8
  \,,\quad
  H_6=-F_8 \,.
\end{eqnarray}
To sub-LO with all intermediate parameters kept in Eq.~(\ref{eq:P-ansatz}),
we have $6-3=3$ sum rules:
\begin{eqnarray}
  \label{eq:sumrule:chi-:subLO}
  &&\,
  C_5=-F_8
  \,,\quad
  H_5=9\,F_7-3\,F_8
  \,,\quad
  H_6=3\,F_9-3\,F_7+3\,F_8
  \,.
\end{eqnarray}

We close this section with  the time-ordered product of a two-body correlation function involving an axial current and a scalar current.
Here we found 12 operators of relevance at leading and subleading orders
\begin{eqnarray}
  \label{temp:O-ansatz:AS-prototype-a0+00-NPA}
  {\cal S}^i_{ab}
  & = &
  \hat{g}_{1}\,d_{ab{\tilde{e}}}\,G^i_{{\tilde{e}}}
  + \hat{g}_{2}\,[T_{a},\,G^i_{b}]_+
  + \hat{g}_{3}\,[T_{b},\,G^i_{a}]_+
  + \hat{g}_{4}\,G^i_{a}\,\delta_{b0}
  + \hat{g}_{5}\,\delta_{ab}\,J^i
  + \hat{g}_{6}\,d_{ab{\tilde{e}}}\,[J^i,\,T_{{\tilde{e}}}]_+
  \nonumber\\ && \,
  +\, \hat{g}_{7}\,[J^i,\,[T_{a},\,T_{b}]_+]_+
  + \hat{g}_{8}\,[J^l,\,[G^i_{a},\,G^l_{b}]_+]_+
  + \hat{g}_{9}\,i\epsilon^{ijk}\,[J^j,\,[T_{a},\,G^k_{b}]_+]_+
  \nonumber\\ && \,
  +\, \hat{g}_{10}\,i\epsilon^{ijk}\,[J^j,\,[T_{b},\,G^k_{a}]_+]_+
  + \hat{g}_{11}\,[J^i,\,T_{a}]_+\,\delta_{b0}
  + \hat{g}_{12}\,i\epsilon^{ijk}\,[J^j,\,G^k_{a}]_+\,\delta_{b0}
  \,,
\end{eqnarray}
where the flavor dummy indices $j$, $k$, and $l$ run from $1$ to $3$ but $\tilde{e}$ does from $0$, $1$ to $8$.
The parameters $\hat{g}_{1-4}$ and $\hat{g}_{5-12}$ are relevant at leading and subleading orders respectively, following the effective scaling of large-$N_c$.
Applying Eqs.~(\ref{result:one-body-operators}) and~(\ref{eq:orthonormalities}) repeatedly is conceptually sufficient to get matrix elements within the effective baryon states.
But the explicit mathematical form of each matrix element is helpful to improve its efficiency.
Those for one- and two-body operators are presented in Ref.~\cite{Lutz:2010se} and for three-body operators in Ref.~\cite{Heo:2019cqo} and in Appendix~\ref{sec:matrix-elements-3}.
When a singlet occurs, $b=0$ and/or $\tilde{e}=0$, the supplementary notation,
\begin{eqnarray}
  &&
  T_0 = \tfrac{1}{\sqrt{6}}\,{\mathbbm 1}
  \,,\quad
  G^i_0 = \tfrac{1}{\sqrt{6}}\,J^i
  \,,\qquad
  \lambda_0 = \sqrt{\tfrac{2}{3}}\,{\bf 1}_{\!3\times 3}
  \,,\quad
  f_{0ab} = 0
  \,,\quad
  d_{0ab} = \sqrt{\tfrac{2}{3}}\,\delta_{ab}
  \,,
\end{eqnarray}
will be useful.
It turns out that the results in Appendix~\ref{sec:matrix-elements-3} played an essential role.
The matching of the matrix elements with those from the chiral Lagrangian is accomplished by using an algebraic computer algorithm.
It leads to the identifications with the correlation functions~(\ref{summary:<08|AS|08>-LHS-c=0-L3+-nre})-(\ref{summary:<10|AS|08>-LHS-c=0-L3+-nre}) as detailed in Tab.~\ref{tab:AS-NPA}.

\begin{table}[t]
  \centering
  \begin{tabular}{R@{\;=\;}L}
{F}_{0}+\frac{3}{2}\,F_4
&-\frac{3}{2}\,\hat{g}_{5}-\hat{g}_{6}-\frac{3}{2}\,\hat{g}_{8}
\\
{F}_{1}
&-\frac{1}{3}\,\hat{g}_{1}-\hat{g}_{3}-\hat{g}_{6}-\frac{2}{3}\,\hat{g}_{8}
\\
{F}_{2}
&-\frac{1}{3}\,\hat{g}_{1}-\hat{g}_{2}-\hat{g}_{6}-\frac{2}{3}\,\hat{g}_{8}
\\
{F}_{3}+\frac{1}{2}\,F_4
&-\frac{1}{3}\,\hat{g}_{1}-\frac{2}{3}\,\hat{g}_{2}-\frac{2}{3}\,\hat{g}_{3}-\frac{1}{2}\,\hat{g}_{5}-\frac{1}{3}\,\hat{g}_{6}-2\,\hat{g}_{7}-\frac{1}{3}\,\hat{g}_{8}
\\
F_5-2\,F_4
&-\frac{2}{3}\,\hat{g}_{1}+2\,\hat{g}_{5}+\frac{4}{3}\,\hat{g}_{6}+\frac{8}{3}\,\hat{g}_{8}
\\
-F_4
&-\frac{1}{3}\,\hat{g}_{1}-\frac{1}{3}\,\hat{g}_{3}-\frac{1}{\sqrt{6}}\,\hat{g}_{4}+\hat{g}_{5}+\frac{2}{3}\,\hat{g}_{6}+\frac{5}{6}\,\hat{g}_{8}
\\
F_6
&\frac{1}{3}\,\hat{g}_{2}-\frac{2}{9}\,\hat{g}_{3}-\frac{1}{3}\,\sqrt{\frac{2}{3}}\,\hat{g}_{4}-\frac{2}{3}\,\hat{g}_{7}+\frac{1}{3}\,\hat{g}_{8}-\sqrt{\frac{2}{3}}\,\hat{g}_{11}
    \\ \hline
{H}_{0}
&-\frac{3}{2}\,\hat{g}_{1}-\frac{9}{2}\,\hat{g}_{2}-\frac{9}{2}\,\hat{g}_{3}-9\,\hat{g}_{6}-27\,\hat{g}_{7}-\frac{39}{4}\,\hat{g}_{8}
\\
{H}_{1}
&0
\\
{H}_{2}
&6\,\hat{g}_{2}+6\,\hat{g}_{3}+36\,\hat{g}_{7}+15\,\hat{g}_{8}
\\
{H}_{3}
&4\,\hat{g}_{6}-3\,\hat{g}_{5}
\\
{H}_{4}
&2\,\hat{g}_{3}-\sqrt{\frac{3}{2}}\,\hat{g}_{4}+12\,\hat{g}_{7}-3\,\sqrt{6}\,\hat{g}_{11}
    \\ \hline
{C}_{0}
&-\hat{g}_{1}-\frac{1}{2}\,\hat{g}_{3}-\frac{1}{2}\,\sqrt{\frac{3}{2}}\,\hat{g}_{4}-\frac{1}{2}\,\hat{g}_{8}+\frac{3}{2}\,\hat{g}_{10}+\frac{3}{2}\,\sqrt{\frac{3}{2}}\,\hat{g}_{12}
\\
{C}_{1}
&\hat{g}_{2}-\frac{5}{6}\,\hat{g}_{3}+\frac{1}{2\,\sqrt{6}}\,\hat{g}_{4}-2\,\hat{g}_{8}-3\,\hat{g}_{9}+\frac{5}{2}\,\hat{g}_{10}-\frac{1}{2}\,\sqrt{\frac{3}{2}}\,\hat{g}_{12}
\\
{C}_{2}
&-\frac{1}{2}\,\hat{g}_{3}-\frac{1}{2}\,\sqrt{\frac{3}{2}}\,\hat{g}_{4}-3\,\hat{g}_{8}+\frac{3}{2}\,\hat{g}_{10}+\frac{3}{2}\,\sqrt{\frac{3}{2}}\,\hat{g}_{12}
\\
{C}_{3}
&-\frac{11}{6}\,\hat{g}_{3}+\frac{1}{2\,\sqrt{6}}\,\hat{g}_{4}-3\,\hat{g}_{8}+\frac{11}{2}\,\hat{g}_{10}-\frac{1}{2}\,\sqrt{\frac{3}{2}}\,\hat{g}_{12}
\\
{C}_{4}
&-2\,\hat{g}_{2}+\frac{1}{3}\,\hat{g}_{3}+\frac{1}{\sqrt{6}}\,\hat{g}_{4}+\hat{g}_{8}+6\,\hat{g}_{9}-\hat{g}_{10}-\sqrt{\frac{3}{2}}\,\hat{g}_{12}
  \end{tabular}
  \caption{
    The low-energy constants introduced in~(\ref{eq:cal:Lag:Q3:chi+}) are correlated with the intermediate parameters of~(\ref{temp:O-ansatz:AS-prototype-a0+00-NPA}) via the large-$N_c$ operator expansion~(\ref{def-largeN-expansion}).
  }
  \label{tab:AS-NPA}
\end{table}

In the $1/N_c$ expansion to leading order (LO),
there are $17-(4)=13$ sum rules as follow.
\begin{eqnarray}
  \label{eq:sumrule:LO:NPA}
  &&
  3\,F_4=- 2\,F_{0}
  \,,\quad
  F_6 = \tfrac{1}{9}\,F_{0}-\tfrac{2}{3}\,F_1-F_2+F_3
  \,,\nonumber\\ &&
  C_{4}=-2\,C_{1}-C_{2}+C_{3}
  \,,\quad
  F_5 = \tfrac{2}{3}\,F_{0}+4\,F_1+4\,F_2-6\,F_3
  \,,\nonumber\\ &&
  2\,C_{0}=5\,F_{0}+6\,F_1+6\,F_2-9\,F_3
  \,,\quad
  6\,C_{1} = F_{0}+12\,F_1-9\,F_3
  \,,\nonumber\\ &&
  2\,C_{2}=-F_{0}-6\,F_1-6\,F_2+9\,F_3
  \,,\quad
  6\,C_{3} = -11\,F_{0}-6\,F_1-18\,F_2+27\,F_3
  \,,\nonumber\\ &&
  2\,H_{0}=- 9\,(F_{0}+F_1+F_2-3\,F_3)
  \,,\quad
  H_{1} = 0
  \,,\quad
  H_{2} = 12\,F_{0}+18\,F_1+18\,F_2-36\,F_3
  \,,\nonumber\\ &&
  H_{3}=0
  \,,\quad
  H_{4} = 2\,F_{0}-3\,F_1
  \,.
\end{eqnarray}
At subleading order (sub-LO) with all 12 operators remained we find $17-(8+4)=5$ sum rules
\begin{eqnarray}
  &&
  \label{eq:sumrule:subLO:NPA}
  2\,C_{0}=-2\,C_{1}+C_{2}+C_{3}-C_{4}+4\,F_{0}+3\,F_5
  \,,\quad
  H_{1} = 0
  \,,\nonumber\\ &&
  2\,H_{0}+3\,H_{2} =-24\,C_{1}-12\,C_{2}+12\,C_{3}-12\,C_{4}+21\,F_{0}+9\,F_1+9\,F_2-27\,F_3+9\,F_5
  \,,\nonumber\\ &&
  4\,H_{3}=9\,F_{0}-3\,F_1-3\,F_2+9\,F_3+12\,F_4+3\,F_5-2\,H_{0}-H_{2}
  \,,\nonumber\\ &&
  H_{4}=\tfrac{3}{2}\,F_{0}+\tfrac{3}{2}\,F_1+\tfrac{15}{2}\,F_2-\tfrac{9}{2}\,F_3+\tfrac{3}{2}\,F_5+9\,F_6
  -H_{0}-\tfrac{1}{2}\,H_{2}
  \,.
\end{eqnarray}

\newpage
\section{Phenomenological tree-level study of the sum rules}
\label{sec:numerics}

We explore the implications of our sum rules. At this stage a phenomenological study is possible only.
In a $\chi$PT framework the accuracy level at which the LEC $F_i$, $C_i$, and $H_i$ enter,
the evaluation of loop contributions to the various correlation functions is indispensable. 
While in principle this can and should be done, a complete study of these contributions is outside the scope of our current study.
Only recently,
a promising novel framework, in which such computations can be performed is available \cite{Lutz:2020dfi,Sauerwein:2021jxb}. The key element is the request to consider loop contributions with on-shell meson and baryon masses. Only then a sufficiently convergent chiral expansion is expected by us in three-flavor $\chi$PT. This request leaves the conventional path of $\chi$PT computations and clearly needs further consolidations. In any case, 
a complete picture is achievable only after further QCD lattice data on matrix elements of axial and pseudoscalar currents became available.

\begin{table}[t]
\setlength{\tabcolsep}{1.9mm}
  \centering
\begin{tabular}{@{}C@{$\,\rightarrow\,$}C@{\,\,\,}|R@{.}L@{$\,\pm$}L|R@{.}LR@{.}LR@{.}L@{}}
  \multicolumn{2}{C|}{\phantom{x}}
  & \multicolumn{3}{c|}{$g_A$}
  & \multicolumn{2}{C|}{\rm Fit \, 1}
  & \multicolumn{2}{C|}{\rm Fit \, 2}
  & \multicolumn{2}{C}{\rm Fit \, 3}
  \\  \hline
  {n}
  & {p}\,{e}^-\bar\nu_e
  &+1&272
  &0.002
&1&244
&1&282
&1&272
  \\
  \Sigma^-
  & \Lambda\,{e}^-\bar\nu_e
  &+0&601
  &0.015
&0&631
&0&579
&0&601
  \\ 
  \Lambda
  & {p}\,{e}^-\bar\nu_e
  &-0&879
  &0.018
&-0&892
&-0&946
&-0&879
  \\
  \Sigma^-
  & {n}\,{e}^-\bar\nu_e
  &+0&340
  &0.017
&0&302
&0&365
&0&340
  \\
  \Xi^-
  & \Lambda\,{e}^-\bar\nu_e
  &+0&306
  &0.061
&0&261
&0&279
&0&306
  \\
  \Xi^0
  & \Sigma^+\,{e}^-\bar\nu_e
  &+1&220
  &0.050
&1&244
&1&139
&1&220
  \\ 
\multicolumn{2}{C|}{\phantom{x}}
  & \multicolumn{3}{c|}{$g_P$}
  & \multicolumn{2}{C|}{\rm Fit \, 1}
  & \multicolumn{2}{C|}{\rm Fit \, 2}
  & \multicolumn{2}{C}{\rm Fit \, 3}
  \\  \hline
    \Delta
  & N\,\pi
  &+2&04
  &0.01
&1&78
&2&00
&2&04
  \\
  \Sigma^*
  & \Lambda\,\pi
  &-1&21
  &0.02
&-1&26
&-1&26
&-1&22
  \\
  \Sigma^*
  & \Sigma\,\pi
  &-0&92
  &0.08
&-1&03
&-0&89
&-0&98
  \\
  \Xi^*
  & \Xi\,\pi
  &-1&00
  &0.03
&-1&26
&-1&02
&-0&99
  \\ \hline
  \multicolumn{5}{C}{\delta_{theory}}
  & \multicolumn{2}{C}{0.145}
  & \multicolumn{2}{C}{0.058}
  & \multicolumn{2}{C}{0.000} \\ \hline
\end{tabular}
  \caption{
    The coupling constants $g_A$ for $\beta$ decays and $g_P$ from Eq.~(\ref{def-gP}) for the  $\pi$ decays are listed. The empirical values  are taken from  \cite{Zyla:2020zbs} but for the $g_A$ in the $\Sigma^- \to \Lambda \,e ^- \bar \nu_e $ decay and all $g_P$ values for which we use ~\cite{Dai:1995zg}. The three fit scenarios with  $\chi^2/{N_{d.o.f.} \simeq 1.0}$ are discussed in the text and are based on the expressions collected in Tab.~\ref{tab:gA:beta:math} and Tab.~\ref{tab:gP:pion:math}. 
  }
\label{tab:gA:gP-fits}
\end{table}

Here we follow the phenomenological path of ~\cite{Dai:1995zg} along which the empirical $\beta$ and pion decay processes of the baryons are analyzed in terms of tree-level expressions.
In our case we use Tab.~\ref{tab:gA:beta:math} and Tab.~\ref{tab:gP:pion:math}.
Corresponding empirical values for the $g_A$ and $g_P$ are collected in Tab.~\ref{tab:gA:gP-fits} alongside with three fit scenarios. 
In our $\chi^2$-square function we consider the empirical uncertainties supplemented by a uniform residual theoretical uncertainty, $\delta_{\rm theory}$, which is added in quadrature. The latter is estimated by the request that $\chi^2/N_{d.o.f} \simeq 1.0$ with $N_{d.o.f} = N_{data} - N_{parameters}$, the conventional number of degrees of freedom in the fit system.

\begin{table}[t]
\setlength{\tabcolsep}{3.1mm}
  \centering
\begin{tabular}{L@{\,\,\,}|R@{.}L@{\,\,\,}R@{.}LR@{.}L}
  & \multicolumn{2}{C|}{\rm Fit\, 1}
  & \multicolumn{2}{C|}{\rm Fit \,2}
  & \multicolumn{2}{C}{\rm Fit\, 3}
  \\ \hline
  F_R
&0&47(12)
& 0&438(40)
&0&485(18) 
  \\
  D_R
&0&77(14)
&0&748(50)
&0&752(14)
  \\
  C_A - \varepsilon_8\,(  \tfrac{2}{3}\, C_0  + C_4 - 2\,F_2 )
&1&78(18)
& 1&86(11)
& 1&637(87)
  \\ \hline
  F_1\hfill [{\rm GeV}^{-2}]
  &0&0
  & -0&15(10)
  & -0&226(50)
  \\
  F_2\hfill [{\rm GeV}^{-2}]
&0&0
&0&22(12)
&-0&081(91)
  \\
  F_3 - \tfrac{1}{3}\,F_0 \hfill[{\rm GeV}^{-2}]
& 0&0
& 0&02(11)
& 0&122(61)
  \\
  F_5 + \tfrac{4}{3}\,F_0 \hfill [{\rm GeV}^{-2}]
& 0&0
&0&175
&0&072(69)
  \\
  C_2 + \tfrac{3}{2}\,C_4 \hfill [{\rm GeV}^{-2}]
&0&0
&0&400
&0&05(13)
  \\ 
  C_3 - \tfrac{1}{2}\,C_4 \hfill [{\rm GeV}^{-2}]
&0&0
  & -0&614
&-0& 710(31)
\end{tabular}
  \caption{
    Low-energy parameters from our fit scenarios in Tab.~\ref{tab:gA:gP-fits}. All fit parameters come with their 1-$\sigma$ error.
    Values without an associated error are derived quantities from our large-$N_c$ sum rules Eq.~(\ref{res-LO-large-NC}).
     }
\label{tab:LEC}
\end{table}

We discuss the fit scenarios.
First we consider the leading order LEC only, which implies that in Tab.~\ref{tab:gA:beta:math} and Tab.~\ref{tab:gP:pion:math} the parameters $F_R, D_R$, and $C_A$ are fitted.
All $F_i$ and $C_i$ are put to zero. Since the $\chi^2/N_{d.o.f} \simeq 1$ by construction the 
decisive result of such a fit is the extracted residual uncertainty of $\delta_{theory} \simeq 0.145$ in this case. The latter value gives an estimate for the size of the higher order effects,
i.e. from the $F_i$ and $C_i$  and the loop contributions, that cannot be absorbed into the leading order LEC.
We note that such type of fits have been done ample times before in the literature (see e.g. \cite{Okun_2014}).

We turn to our second fit scenario, in which we insist on the leading order sum rules: Eqs.~(\ref{eq:sumrule:chi-:LO}) and (\ref{eq:sumrule:LO:NPA}).
Besides the parameters $F_R, D_R$, and $C_A$ that adds 4 additional fit parameters $F_{0-3}$ only.
We note that according to Eq.~(\ref{def-CA}) and Eq.~(\ref{eq:sumrule:chi-:LO}) the dependence on the parameter $C_5$ was fully absorbed into the definition of our $C_A$ parameter. A detailed inspection of Tab.~\ref{tab:gA:beta:math} reveals that, 
under the leading order large-$N_c$ sum rules, 
the considered $\beta$ decay processes  are sensitive to the particular parameter combination $F_3-F_0/3$ only. In turn the parameters $F_0$ and $F_3$ cannot be determined independently and therefore our second scenario has 6 fit parameters only. Note here that the parameter $F_4$ does not enter any of the expressions in Tab.~\ref{tab:gA:beta:math}. Again we determined 
the residual uncertainty, with $\delta_{theory} \simeq 0.058$ in this case, by the condition that 
$\chi^2/N_{d.o.f} \simeq 1.0$. It is reassuring to see a significant drop of $\delta_{theory}$ as we include correction effects predicted by our large-$N_c$ sum rules. 

Consider our third fit scenario,
in which we insist on the subleading order sum rules: Eqs.~(\ref{eq:sumrule:chi-:subLO}) and (\ref{eq:sumrule:subLO:NPA}). Initially, there would be 15 parameters 
relevant for our phenomenological study. Clearly, there must be a set of particular combinations that can be determined only. Besides the $D_R, F_R$ and $F_1, F_2$ parameters we identified the following 5 combinations
\begin{eqnarray}
&& C_A -\varepsilon_8\,(  \tfrac{2}{3}\, C_0  + C_4 - 2\,F_2 ) \,,
\nonumber\\
&& F_3 -  \tfrac{1}{3}\,F_0, \quad  F_5 +  \tfrac{4}{3}\,F_0 ,\quad  C_2 + \tfrac{3}{2}\,C_4 ,\quad  C_3 - \tfrac{1}{2} \,C_4 \,,
\end{eqnarray}
which are determined in our fit. Note that the parameters $F_4$ and $F_6 $ do not enter Tab.~\ref{tab:gA:beta:math}. While $F_6$ enters in the definition of $F_R$ in Eq.~(\ref{def-FR-DR}), 
the parameters $F_4$ is unconstrained. Similarly the parameter $C_1$ does not enter Tab.~\ref{tab:gP:pion:math} and remains unconstrained here. 
Again we find a further significant drop in the 
residual uncertainty with $\delta_{theory} \simeq 0.000 $ in this case.

Our sets of LEC from the three fit scenarios are collected in Tab.~\ref{tab:LEC}. 
First we observe that our $F_R$, $D_R$, and $C_A-\varepsilon_8\,(  \tfrac{2}{3}\, C_0  + C_4 - 2\,F_2 ) $ for every fit are quite consistent with
$F\sim 0.487\pm 0.002$, $D\sim 0.743\pm 0.002$, and $C\sim 1.487\pm 0.005$ from~\cite{Lutz:2018cqo,Guo:2019nyp} and even roughly compatible with
$F\sim 0.45$, $D\sim 0.80$, and $C\sim 1.6$ from~\cite{Okun_2014,Bernard:2003xf} as well.

As we emphasized before the role of the subleading chiral counter terms $F_i$, $C_i$, and $H_i$ in our 
tree-level study can be seen only at a phenomenological level, as long as the mandatory loop 
contributions are not yet worked in.
Still
we find it interesting that such an analysis seems to be working unexpectedly well.
A comparison of Fit 3 LEC with those of the Fit 2 scenario reveals 
a not too disastrous pattern as we include higher order terms in the $1/N_c$ expansion. The values 
without an associated 1-$\sigma$ error in Tab.~\ref{tab:LEC} follow with
\begin{eqnarray}
&&  \tfrac{2}{3}\, C_0  + C_4 - 2\,F_2  = 0 \,,
\nonumber\\
&&  F_5 +  \tfrac{4}{3}\,F_0  = 4\, F_1 + 4\, F_2 - 2\,( 3\,F_3 - F_0) \,, \qquad 
\nonumber\\
&&  C_2 + \tfrac{3}{2}\,C_4 = -3\, (2\, F_1+F_2- 3 \,F_3 + F_0)\,, 
\nonumber\\
&& C_3 - \tfrac{1}{2} \,C_4 = -3 \,F_2+ 3\,F_3 - F_0 \,,
\label{res-LO-large-NC}
\end{eqnarray}
from the other LEC in that fit.

\newpage
\section{Summary}
In this work we completed our study of the chiral SU(3) Lagrangian 
with baryon octet and decuplet fields at chiral order $Q^3$. 
The $23$ terms which involve the symmetry-breaking fields were scrutinized in the context of large-$N_c$ 
QCD. Sets of sum rules were established. At leading order in the $1/N_c$ expansion we find $18$ conditions that imply a significant parameter reduction. At subleading order, the number of sum rules comes at $8$.   

A phenomenological tree-level application to the empirical axial-vector coupling constants of the baryon octet and strong decay channels of the baryon decuplet states was presented. While the empirical values can be well reproduced, it is premature to draw any significant conclusions. It remains to 
supplement such results by loop corrections.  A novel framework for the latter was recently developed in 
Refs.~\cite{Lutz:2020dfi,Sauerwein:2021jxb}, where particular emphasis was put on the importance to use on-shell masses for mesons and baryons in loop contributions from the chiral Lagrangian. 

With the result of this work, we are well prepared to analyze QCD lattice simulation data to come for axial-vector form factors of the baryon octet and decuplet states.

\vskip0.3cm
{\bfseries{Acknowledgments}}
\vskip0.3cm
Y. Heo and C. Kobdaj acknowledge support from Suranaree University of Technology (SUT),
Thailand Science Research and Innovation (TSRI) and SUT-CHE-NRU (Grant No. Ft61/15/2020).

\newpage
\appendix
\section{Matrix elements for 3-body operators}
\label{sec:matrix-elements-3}
The flavor index $a$ coming up in this section is restricted to octet components only, i.e. $1$ to $8$.
\begin{eqnarray}
  ({d},\bar\chi|\,[J^h,[G^i_a,J^j]_+]_+\,|c,\chi)
  & = &
  \delta^{ij}\,\sigma^h_{\bar\chi\chi}\,(
  \tfrac{1}{2}\, d_{ac{d}} + \tfrac{1}{3}\,if_{ac{d}}
  )
  \,,\nonumber\\
  ({d},\bar\chi|\,[J^h,[J^i,J^j]_+]_+\,|c,\chi)
  & = &
  \tfrac{1}{2}\,\delta^{ij}\,\sigma^h_{\bar\chi\chi}\,\delta_{{d}c}
  \,,\\ \nonumber\\
  (nop,\bar\chi|\,[J^h,[G^i_a,J^j]_+]_+\,|klm,\chi)
  & = &
  - \tfrac{3}{8}\,(
  2\,S^i\,\sigma^h\,S^{j\dagger}
  + 2\,S^j\,\sigma^h\,S^{i\dagger}
  + S^h\,\sigma^i\,S^{j\dagger}
  + S^h\,\sigma^j\,S^{i\dagger}
  \nonumber\\ && \,\quad
  +\, S^{i}\,\sigma^j\,S^{h\dagger}
  + S^{j}\,\sigma^i\,S^{h\dagger}
  - 8\,\delta^{ij}\,S^r\,\sigma^h\,S^{r\dagger}
  \nonumber\\ && \,\quad
  -\, 5\,\delta^{hj}\,S^r\,\sigma^i\,S^{r\dagger}
  - 5\,\delta^{hi}\,S^r\,\sigma^j\,S^{r\dagger}
  )_{\bar\chi\chi}
  \,\delta^{nop}_{xyz}\,\Lambda^{a,xyz}_{klm}
  \,,\nonumber\\
  (nop,\bar\chi|\,[J^h,[J^i,J^j]_+]_+\,|klm,\chi)
  & = &
  - \tfrac{3}{4}\,(
  2\,S^i\,\sigma^h\,S^{j\dagger}
  + 2\,S^j\,\sigma^h\,S^{i\dagger}
  + S^h\,\sigma^i\,S^{j\dagger}
  + S^h\,\sigma^j\,S^{i\dagger}
  \nonumber\\ && \,\quad
  +\, S^{i}\,\sigma^j\,S^{h\dagger}
  + S^{j}\,\sigma^i\,S^{h\dagger}
  - 8\,\delta^{ij}\,S^r\,\sigma^h\,S^{r\dagger}
  \nonumber\\ && \,\quad
  -\, 5\,\delta^{hj}\,S^r\,\sigma^i\,S^{r\dagger}
  - 5\,\delta^{hi}\,S^r\,\sigma^j\,S^{r\dagger}
  )_{\bar\chi\chi}
  \,\delta^{nop}_{klm}
  \,,\\ \nonumber\\
  (nop,\bar\chi|\,[J^h,[G^i_a,J^j]_+]_+\,|c,\chi)
  & = &
  \tfrac{1}{8\,\sqrt{2}}\,(
  -\, 2\,\delta^{ij}\,S^h
  - 3\,i\epsilon^{ijr}\,[
  S^{r}\,\sigma^{h} + S^{h}\,\sigma^{r}
  ]
  \nonumber\\ && \, \qquad
  +\, 12\,\delta^{jh}\,S^i
  - i\epsilon^{jhr}\,[
  S^{r}\,\sigma^{i} + S^{i}\,\sigma^{r}
  ]
  \nonumber\\ && \, \qquad
  -\, 6\,\delta^{hi}\,S^j
  + i\epsilon^{hir}\,[
  S^{r}\,\sigma^{j} + S^{j}\,\sigma^{r}
  ]
  )_{\bar\chi\chi}\,\Lambda^{nop}_{ac}
  \,,\nonumber\\
  (nop,\bar\chi|\,[J^h,[J^i,J^j]_+]_+\,|c,\chi)
  & = &
  0
  \,.
\end{eqnarray}

Their particular choices are
\begin{eqnarray}
  ({d},\bar\chi|\,[J^l,[G^i_a,J^l]_+]_+\,|c,\chi)
  & = &
  \sigma^i_{\bar\chi\chi}\,(
  \tfrac{1}{2}\, d_{ac{d}} + \tfrac{1}{3}\,if_{ac{d}}
  )
  \,,\nonumber\\
  ({d},\bar\chi|\,[J^l,[J^i,G^l_a]_+]_+\,|c,\chi)
  & = &
  \delta_{hj}\,({d},\bar\chi|\,[J^h,[G^j_a,J^i]_+]_+\,|c,\chi)
  =
  \sigma^i_{\bar\chi\chi}\,(
  \tfrac{1}{2}\, d_{ac{d}} + \tfrac{1}{3}\,if_{ac{d}}
  )
  \,,\nonumber\\
  ({d},\bar\chi|\,[J^l,[J^i,J^l]_+]_+\,|c,\chi)
  & = &
  \tfrac{1}{2}\,\sigma^i_{\bar\chi\chi}\,\delta_{{d}c}
  \,,\nonumber\\
  ({d},\bar\chi|\,[J^l,[J^l,G^i_a]_+]_+\,|c,\chi)
  & = &
  \sigma^i_{\bar\chi\chi}\,(
  \tfrac{1}{2}\, d_{ac{d}} + \tfrac{1}{3}\,if_{ac{d}}
  )
  \,,\nonumber\\
  ({d},\bar\chi|\,[J^i,[J^l,G^l_a]_+]_+\,|c,\chi)
  & = &
  3\,\sigma^i_{\bar\chi\chi}\,(
  \tfrac{1}{2}\, d_{ac{d}} + \tfrac{1}{3}\,if_{ac{d}}
  )
  \,,\\ \nonumber\\
  (nop,\bar\chi|\,[J^r,[G^i_a,J^r]_+]_+\,|klm,\chi)
  & = &
  \tfrac{39}{4}\,(S^r\,\sigma^i\,S^{r\dagger})_{\bar\chi\chi}
  \,\delta^{nop}_{xyz}\,\Lambda^{a,xyz}_{klm}
  \,,\nonumber\\
  (nop,\bar\chi|\,[J^r,[J^i,G^r_a]_+]_+\,|klm,\chi)
  & = &
  \tfrac{39}{4}\,(S^r\,\sigma^i\,S^{r\dagger})_{\bar\chi\chi}
  \,\delta^{nop}_{xyz}\,\Lambda^{a,xyz}_{klm}
  \,,\nonumber\\
  (nop,\bar\chi|\,[J^r,[J^i,J^r]_+]_+\,|klm,\chi)
  & = &
  \tfrac{39}{2}\,(S^r\,\sigma^i\,S^{r\dagger})_{\bar\chi\chi}
  \,\delta^{nop}_{klm}
  \,,\nonumber\\
  (nop,\bar\chi|\,[J^r,[J^r,G^i_a]_+]_+\,|klm,\chi)
  & = &
  \tfrac{39}{4}\,(S^r\,\sigma^i\,S^{r\dagger})_{\bar\chi\chi}
  \,\delta^{nop}_{xyz}\,\Lambda^{a,xyz}_{klm}
  \,,\nonumber\\
  (nop,\bar\chi|\,[J^i,[J^r,G^r_a]_+]_+\,|klm,\chi)
  & = &
  \tfrac{45}{4}\,(S^r\,\sigma^i\,S^{r\dagger})_{\bar\chi\chi}
  \,\delta^{nop}_{xyz}\,\Lambda^{a,xyz}_{klm}
  \,,\\ \nonumber\\
  (nop,\bar\chi|\,[J^l,[G^i_a,J^l]_+]_+\,|c,\chi)
  & = &
  \tfrac{7}{2\,\sqrt{2}}\,S^i_{\bar\chi\chi}\,\Lambda^{nop}_{ac}
  \,,\nonumber\\
  (nop,\bar\chi|\,[J^l,[J^i,G^l_a]_+]_+\,|c,\chi)
  & = &
  - \tfrac{1}{\sqrt{2}}\,S^i_{\bar\chi\chi}\,\Lambda^{nop}_{ac}
  \,,\nonumber\\
  (nop,\bar\chi|\,[J^l,[J^i,J^l]_+]_+\,|c,\chi)
  & = &
  0
  \,,\nonumber\\
  (nop,\bar\chi|\,[J^l,[J^l,G^i_a]_+]_+\,|c,\chi)
  & = &
  \tfrac{7}{2\,\sqrt{2}}\,S^i_{\bar\chi\chi}\,\Lambda^{nop}_{ac}
  \,,\nonumber\\
  (nop,\bar\chi|\,[J^i,[J^l,G^l_a]_+]_+\,|c,\chi)
  & = &
  0
  \,.
\end{eqnarray}


\bibliography{bibliography_full}

\begin{thebibliography}{46}%
\makeatletter
\providecommand \@ifxundefined [1]{%
 \@ifx{#1\undefined}
}%
\providecommand \@ifnum [1]{%
 \ifnum #1\expandafter \@firstoftwo
 \else \expandafter \@secondoftwo
 \fi
}%
\providecommand \@ifx [1]{%
 \ifx #1\expandafter \@firstoftwo
 \else \expandafter \@secondoftwo
 \fi
}%
\providecommand \natexlab [1]{#1}%
\providecommand \enquote  [1]{``#1''}%
\providecommand \bibnamefont  [1]{#1}%
\providecommand \bibfnamefont [1]{#1}%
\providecommand \citenamefont [1]{#1}%
\providecommand \href@noop [0]{\@secondoftwo}%
\providecommand \href [0]{\begingroup \@sanitize@url \@href}%
\providecommand \@href[1]{\@@startlink{#1}\@@href}%
\providecommand \@@href[1]{\endgroup#1\@@endlink}%
\providecommand \@sanitize@url [0]{\catcode `\\12\catcode `\$12\catcode
  `\&12\catcode `\#12\catcode `\^12\catcode `\_12\catcode `\%12\relax}%
\providecommand \@@startlink[1]{}%
\providecommand \@@endlink[0]{}%
\providecommand \url  [0]{\begingroup\@sanitize@url \@url }%
\providecommand \@url [1]{\endgroup\@href {#1}{\urlprefix }}%
\providecommand \urlprefix  [0]{URL }%
\providecommand \Eprint [0]{\href }%
\providecommand \doibase [0]{http://dx.doi.org/}%
\providecommand \selectlanguage [0]{\@gobble}%
\providecommand \bibinfo  [0]{\@secondoftwo}%
\providecommand \bibfield  [0]{\@secondoftwo}%
\providecommand \translation [1]{[#1]}%
\providecommand \BibitemOpen [0]{}%
\providecommand \bibitemStop [0]{}%
\providecommand \bibitemNoStop [0]{.\EOS\space}%
\providecommand \EOS [0]{\spacefactor3000\relax}%
\providecommand \BibitemShut  [1]{\csname bibitem#1\endcsname}%
\let\auto@bib@innerbib\@empty
\bibitem [{\citenamefont {Weinberg}(1979)}]{Weinberg:1979poi}%
  \BibitemOpen
  \bibfield  {author} {\bibinfo {author} {\bibfnamefont {S.}~\bibnamefont
  {Weinberg}},\ }\href@noop {} {\bibfield  {journal} {\bibinfo  {journal}
  {Physica A: Statistical Mechanics and its Applications}\ }\textbf {\bibinfo
  {volume} {96}},\ \bibinfo {pages} {327 } (\bibinfo {year}
  {1979})}\BibitemShut {NoStop}%
\bibitem [{\citenamefont {Gasser}\ and\ \citenamefont
  {Leutwyler}(1985)}]{Gasser:1984gg}%
  \BibitemOpen
  \bibfield  {author} {\bibinfo {author} {\bibfnamefont {J.}~\bibnamefont
  {Gasser}}\ and\ \bibinfo {author} {\bibfnamefont {H.}~\bibnamefont
  {Leutwyler}},\ }\href@noop {} {\bibfield  {journal} {\bibinfo  {journal}
  {Nucl.Phys.}\ }\textbf {\bibinfo {volume} {B250}},\ \bibinfo {pages} {465}
  (\bibinfo {year} {1985})}\BibitemShut {NoStop}%
\bibitem [{\citenamefont {Krause}(1990)}]{Krause:1990xc}%
  \BibitemOpen
  \bibfield  {author} {\bibinfo {author} {\bibfnamefont {A.}~\bibnamefont
  {Krause}},\ }\href@noop {} {\bibfield  {journal} {\bibinfo  {journal} {Helv.
  Phys. Acta}\ }\textbf {\bibinfo {volume} {63}},\ \bibinfo {pages} {3}
  (\bibinfo {year} {1990})}\BibitemShut {NoStop}%
\bibitem [{\citenamefont {Bernard}\ \emph {et~al.}(1993)\citenamefont
  {Bernard}, \citenamefont {Kaiser},\ and\ \citenamefont
  {Meissner}}]{Bernard:1993nj}%
  \BibitemOpen
  \bibfield  {author} {\bibinfo {author} {\bibfnamefont {V.}~\bibnamefont
  {Bernard}}, \bibinfo {author} {\bibfnamefont {N.}~\bibnamefont {Kaiser}}, \
  and\ \bibinfo {author} {\bibfnamefont {U.~G.}\ \bibnamefont {Meissner}},\
  }\href@noop {} {\bibfield  {journal} {\bibinfo  {journal} {Z.Phys.}\ }\textbf
  {\bibinfo {volume} {C60}},\ \bibinfo {pages} {111} (\bibinfo {year}
  {1993})}\BibitemShut {NoStop}%
\bibitem [{\citenamefont {Kaiser}\ \emph {et~al.}(1997)\citenamefont {Kaiser},
  \citenamefont {Waas},\ and\ \citenamefont {Weise}}]{Kaiser:1996js}%
  \BibitemOpen
  \bibfield  {author} {\bibinfo {author} {\bibfnamefont {N.}~\bibnamefont
  {Kaiser}}, \bibinfo {author} {\bibfnamefont {T.}~\bibnamefont {Waas}}, \ and\
  \bibinfo {author} {\bibfnamefont {W.}~\bibnamefont {Weise}},\ }\href@noop {}
  {\bibfield  {journal} {\bibinfo  {journal} {Nucl.Phys.}\ }\textbf {\bibinfo
  {volume} {A612}},\ \bibinfo {pages} {297} (\bibinfo {year}
  {1997})}\BibitemShut {NoStop}%
\bibitem [{\citenamefont {Fearing}\ and\ \citenamefont
  {Scherer}(2000)}]{Fearing:1999fw}%
  \BibitemOpen
  \bibfield  {author} {\bibinfo {author} {\bibfnamefont {H.}~\bibnamefont
  {Fearing}}\ and\ \bibinfo {author} {\bibfnamefont {S.}~\bibnamefont
  {Scherer}},\ }\href@noop {} {\bibfield  {journal} {\bibinfo  {journal}
  {Phys.Rev.}\ }\textbf {\bibinfo {volume} {C62}},\ \bibinfo {pages} {034003}
  (\bibinfo {year} {2000})}\BibitemShut {NoStop}%
\bibitem [{\citenamefont {Lutz}\ and\ \citenamefont
  {Kolomeitsev}(2002)}]{Lutz:2001yb}%
  \BibitemOpen
  \bibfield  {author} {\bibinfo {author} {\bibfnamefont {M.}~\bibnamefont
  {Lutz}}\ and\ \bibinfo {author} {\bibfnamefont {E.}~\bibnamefont
  {Kolomeitsev}},\ }\href@noop {} {\bibfield  {journal} {\bibinfo  {journal}
  {Nucl.Phys.}\ }\textbf {\bibinfo {volume} {A700}},\ \bibinfo {pages} {193}
  (\bibinfo {year} {2002})}\BibitemShut {NoStop}%
\bibitem [{\citenamefont {Kolomeitsev}\ and\ \citenamefont
  {Lutz}(2004)}]{Kolomeitsev:2003kt}%
  \BibitemOpen
  \bibfield  {author} {\bibinfo {author} {\bibfnamefont {E.}~\bibnamefont
  {Kolomeitsev}}\ and\ \bibinfo {author} {\bibfnamefont {M.}~\bibnamefont
  {Lutz}},\ }\href@noop {} {\bibfield  {journal} {\bibinfo  {journal}
  {Phys.Lett.}\ }\textbf {\bibinfo {volume} {B585}},\ \bibinfo {pages} {243}
  (\bibinfo {year} {2004})}\BibitemShut {NoStop}%
\bibitem [{\citenamefont {Oller}\ \emph {et~al.}(2007)\citenamefont {Oller},
  \citenamefont {Verbeni},\ and\ \citenamefont {Prades}}]{Oller:2007qd}%
  \BibitemOpen
  \bibfield  {author} {\bibinfo {author} {\bibfnamefont {J.~A.}\ \bibnamefont
  {Oller}}, \bibinfo {author} {\bibfnamefont {M.}~\bibnamefont {Verbeni}}, \
  and\ \bibinfo {author} {\bibfnamefont {J.}~\bibnamefont {Prades}},\
  }\href@noop {} {\  (\bibinfo {year} {2007})}\BibitemShut {NoStop}%
\bibitem [{\citenamefont {Caro~Ramon}\ \emph {et~al.}(2000)\citenamefont
  {Caro~Ramon}, \citenamefont {Kaiser}, \citenamefont {Wetzel},\ and\
  \citenamefont {Weise}}]{CaroRamon:1999jf}%
  \BibitemOpen
  \bibfield  {author} {\bibinfo {author} {\bibfnamefont {J.}~\bibnamefont
  {Caro~Ramon}}, \bibinfo {author} {\bibfnamefont {N.}~\bibnamefont {Kaiser}},
  \bibinfo {author} {\bibfnamefont {S.}~\bibnamefont {Wetzel}}, \ and\ \bibinfo
  {author} {\bibfnamefont {W.}~\bibnamefont {Weise}},\ }\href@noop {}
  {\bibfield  {journal} {\bibinfo  {journal} {Nucl.Phys.}\ }\textbf {\bibinfo
  {volume} {A672}},\ \bibinfo {pages} {249} (\bibinfo {year}
  {2000})}\BibitemShut {NoStop}%
\bibitem [{\citenamefont {Fernando}\ and\ \citenamefont
  {Goity}(2018)}]{Fernando:2017yqd}%
  \BibitemOpen
  \bibfield  {author} {\bibinfo {author} {\bibfnamefont {I.~P.}\ \bibnamefont
  {Fernando}}\ and\ \bibinfo {author} {\bibfnamefont {J.~L.}\ \bibnamefont
  {Goity}},\ }\href@noop {} {\bibfield  {journal} {\bibinfo  {journal} {Phys.
  Rev. D}\ }\textbf {\bibinfo {volume} {97}},\ \bibinfo {pages} {054010}
  (\bibinfo {year} {2018})}\BibitemShut {NoStop}%
\bibitem [{\citenamefont {Fernando}\ and\ \citenamefont
  {Goity}(2020)}]{Fernando:2019upo}%
  \BibitemOpen
  \bibfield  {author} {\bibinfo {author} {\bibfnamefont {I.~P.}\ \bibnamefont
  {Fernando}}\ and\ \bibinfo {author} {\bibfnamefont {J.~L.}\ \bibnamefont
  {Goity}},\ }\href@noop {} {\bibfield  {journal} {\bibinfo  {journal} {Phys.
  Rev. D}\ }\textbf {\bibinfo {volume} {101}},\ \bibinfo {pages} {054026}
  (\bibinfo {year} {2020})}\BibitemShut {NoStop}%
\bibitem [{\citenamefont {Heo}\ \emph {et~al.}(2019)\citenamefont {Heo},
  \citenamefont {Kobdaj},\ and\ \citenamefont {Lutz}}]{Heo:2019cqo}%
  \BibitemOpen
  \bibfield  {author} {\bibinfo {author} {\bibfnamefont {Y.}~\bibnamefont
  {Heo}}, \bibinfo {author} {\bibfnamefont {C.}~\bibnamefont {Kobdaj}}, \ and\
  \bibinfo {author} {\bibfnamefont {M.~F.~M.}\ \bibnamefont {Lutz}},\
  }\href@noop {} {\bibfield  {journal} {\bibinfo  {journal} {Phys. Rev.}\
  }\textbf {\bibinfo {volume} {D100}},\ \bibinfo {pages} {094035} (\bibinfo
  {year} {2019})}\BibitemShut {NoStop}%
\bibitem [{\citenamefont {Holmberg}\ and\ \citenamefont
  {Leupold}(2019)}]{Holmberg:2019ltw}%
  \BibitemOpen
  \bibfield  {author} {\bibinfo {author} {\bibfnamefont {M.}~\bibnamefont
  {Holmberg}}\ and\ \bibinfo {author} {\bibfnamefont {S.}~\bibnamefont
  {Leupold}},\ }\href@noop {} {\bibfield  {journal} {\bibinfo  {journal} {Phys.
  Rev. D}\ }\textbf {\bibinfo {volume} {100}},\ \bibinfo {pages} {114001}
  (\bibinfo {year} {2019})}\BibitemShut {NoStop}%
\bibitem [{\citenamefont {Sauerwein}\ \emph {et~al.}(2021)\citenamefont
  {Sauerwein}, \citenamefont {Lutz},\ and\ \citenamefont
  {Timmermans}}]{Sauerwein:2021jxb}%
  \BibitemOpen
  \bibfield  {author} {\bibinfo {author} {\bibfnamefont {U.}~\bibnamefont
  {Sauerwein}}, \bibinfo {author} {\bibfnamefont {M.~F.~M.}\ \bibnamefont
  {Lutz}}, \ and\ \bibinfo {author} {\bibfnamefont {R.~G.~E.}\ \bibnamefont
  {Timmermans}},\ }\href@noop {} {\  (\bibinfo {year} {2021})}\BibitemShut
  {NoStop}%
\bibitem [{\citenamefont {'t~Hooft}(1974)}]{Hooft:1973jz}%
  \BibitemOpen
  \bibfield  {author} {\bibinfo {author} {\bibfnamefont {G.}~\bibnamefont
  {'t~Hooft}},\ }\href@noop {} {\bibfield  {journal} {\bibinfo  {journal}
  {Nucl.Phys.}\ }\textbf {\bibinfo {volume} {B72}},\ \bibinfo {pages} {461}
  (\bibinfo {year} {1974})}\BibitemShut {NoStop}%
\bibitem [{\citenamefont {Witten}(1979)}]{Witten:1979kh}%
  \BibitemOpen
  \bibfield  {author} {\bibinfo {author} {\bibfnamefont {E.}~\bibnamefont
  {Witten}},\ }\href@noop {} {\bibfield  {journal} {\bibinfo  {journal}
  {Nucl.Phys.}\ }\textbf {\bibinfo {volume} {B160}},\ \bibinfo {pages} {57}
  (\bibinfo {year} {1979})}\BibitemShut {NoStop}%
\bibitem [{\citenamefont {Luty}\ and\ \citenamefont
  {March-Russell}(1994)}]{Luty:1993fu}%
  \BibitemOpen
  \bibfield  {author} {\bibinfo {author} {\bibfnamefont {M.~A.}\ \bibnamefont
  {Luty}}\ and\ \bibinfo {author} {\bibfnamefont {J.}~\bibnamefont
  {March-Russell}},\ }\href@noop {} {\bibfield  {journal} {\bibinfo  {journal}
  {Nucl.Phys.}\ }\textbf {\bibinfo {volume} {B426}},\ \bibinfo {pages} {71}
  (\bibinfo {year} {1994})}\BibitemShut {NoStop}%
\bibitem [{\citenamefont {Dashen}\ \emph {et~al.}(1995)\citenamefont {Dashen},
  \citenamefont {Jenkins},\ and\ \citenamefont {Manohar}}]{Dashen:1994qi}%
  \BibitemOpen
  \bibfield  {author} {\bibinfo {author} {\bibfnamefont {R.~F.}\ \bibnamefont
  {Dashen}}, \bibinfo {author} {\bibfnamefont {E.~E.}\ \bibnamefont {Jenkins}},
  \ and\ \bibinfo {author} {\bibfnamefont {A.~V.}\ \bibnamefont {Manohar}},\
  }\href@noop {} {\bibfield  {journal} {\bibinfo  {journal} {Phys.Rev.}\
  }\textbf {\bibinfo {volume} {D51}},\ \bibinfo {pages} {3697} (\bibinfo {year}
  {1995})}\BibitemShut {NoStop}%
\bibitem [{\citenamefont {Lutz}\ and\ \citenamefont
  {Semke}(2011)}]{Lutz:2010se}%
  \BibitemOpen
  \bibfield  {author} {\bibinfo {author} {\bibfnamefont {M.}~\bibnamefont
  {Lutz}}\ and\ \bibinfo {author} {\bibfnamefont {A.}~\bibnamefont {Semke}},\
  }\href@noop {} {\bibfield  {journal} {\bibinfo  {journal} {Phys.Rev.}\
  }\textbf {\bibinfo {volume} {D83}},\ \bibinfo {pages} {034008} (\bibinfo
  {year} {2011})}\BibitemShut {NoStop}%
\bibitem [{\citenamefont {Lutz}\ \emph {et~al.}(2014)\citenamefont {Lutz},
  \citenamefont {Bavontaweepanya}, \citenamefont {Kobdaj},\ and\ \citenamefont
  {Schwarz}}]{Lutz:2014oxa}%
  \BibitemOpen
  \bibfield  {author} {\bibinfo {author} {\bibfnamefont {M.~F.~M.}\
  \bibnamefont {Lutz}}, \bibinfo {author} {\bibfnamefont {R.}~\bibnamefont
  {Bavontaweepanya}}, \bibinfo {author} {\bibfnamefont {C.}~\bibnamefont
  {Kobdaj}}, \ and\ \bibinfo {author} {\bibfnamefont {K.}~\bibnamefont
  {Schwarz}},\ }\href@noop {} {\bibfield  {journal} {\bibinfo  {journal} {Phys.
  Rev.}\ }\textbf {\bibinfo {volume} {D90}},\ \bibinfo {pages} {054505}
  (\bibinfo {year} {2014})}\BibitemShut {NoStop}%
\bibitem [{\citenamefont {Lutz}\ \emph {et~al.}(2018)\citenamefont {Lutz},
  \citenamefont {Heo},\ and\ \citenamefont {Guo}}]{Lutz:2018cqo}%
  \BibitemOpen
  \bibfield  {author} {\bibinfo {author} {\bibfnamefont {M.~F.~M.}\
  \bibnamefont {Lutz}}, \bibinfo {author} {\bibfnamefont {Y.}~\bibnamefont
  {Heo}}, \ and\ \bibinfo {author} {\bibfnamefont {X.-Y.}\ \bibnamefont
  {Guo}},\ }\href@noop {} {\bibfield  {journal} {\bibinfo  {journal} {Nucl.
  Phys.}\ }\textbf {\bibinfo {volume} {A977}},\ \bibinfo {pages} {146}
  (\bibinfo {year} {2018})}\BibitemShut {NoStop}%
\bibitem [{\citenamefont {Aubin}\ \emph {et~al.}(2004)\citenamefont {Aubin},
  \citenamefont {Bernard}, \citenamefont {DeTar}, \citenamefont {Osborn},
  \citenamefont {Gottlieb} \emph {et~al.}}]{Aubin:2004wf}%
  \BibitemOpen
  \bibfield  {author} {\bibinfo {author} {\bibfnamefont {C.}~\bibnamefont
  {Aubin}}, \bibinfo {author} {\bibfnamefont {C.}~\bibnamefont {Bernard}},
  \bibinfo {author} {\bibfnamefont {C.}~\bibnamefont {DeTar}}, \bibinfo
  {author} {\bibfnamefont {J.}~\bibnamefont {Osborn}}, \bibinfo {author}
  {\bibfnamefont {S.}~\bibnamefont {Gottlieb}},  \emph {et~al.},\ }\href@noop
  {} {\bibfield  {journal} {\bibinfo  {journal} {Phys.Rev.}\ }\textbf {\bibinfo
  {volume} {D70}},\ \bibinfo {pages} {094505} (\bibinfo {year}
  {2004})}\BibitemShut {NoStop}%
\bibitem [{\citenamefont {Walker-Loud}\ \emph {et~al.}(2009)\citenamefont
  {Walker-Loud}, \citenamefont {Lin}, \citenamefont {Richards}, \citenamefont
  {Edwards}, \citenamefont {Engelhardt} \emph {et~al.}}]{WalkerLoud:2008bp}%
  \BibitemOpen
  \bibfield  {author} {\bibinfo {author} {\bibfnamefont {A.}~\bibnamefont
  {Walker-Loud}}, \bibinfo {author} {\bibfnamefont {H.-W.}\ \bibnamefont
  {Lin}}, \bibinfo {author} {\bibfnamefont {D.}~\bibnamefont {Richards}},
  \bibinfo {author} {\bibfnamefont {R.}~\bibnamefont {Edwards}}, \bibinfo
  {author} {\bibfnamefont {M.}~\bibnamefont {Engelhardt}},  \emph {et~al.},\
  }\href@noop {} {\bibfield  {journal} {\bibinfo  {journal} {Phys.Rev.}\
  }\textbf {\bibinfo {volume} {D79}},\ \bibinfo {pages} {054502} (\bibinfo
  {year} {2009})}\BibitemShut {NoStop}%
\bibitem [{\citenamefont {Aoki}\ \emph {et~al.}(2009)\citenamefont {Aoki} \emph
  {et~al.}}]{Aoki:2008sm}%
  \BibitemOpen
  \bibfield  {author} {\bibinfo {author} {\bibfnamefont {S.}~\bibnamefont
  {Aoki}} \emph {et~al.},\ }\href@noop {} {\bibfield  {journal} {\bibinfo
  {journal} {Phys.Rev.}\ }\textbf {\bibinfo {volume} {D79}},\ \bibinfo {pages}
  {034503} (\bibinfo {year} {2009})}\BibitemShut {NoStop}%
\bibitem [{\citenamefont {Lin}\ \emph {et~al.}(2009)\citenamefont {Lin} \emph
  {et~al.}}]{Lin:2008pr}%
  \BibitemOpen
  \bibfield  {author} {\bibinfo {author} {\bibfnamefont {H.-W.}\ \bibnamefont
  {Lin}} \emph {et~al.},\ }\href@noop {} {\bibfield  {journal} {\bibinfo
  {journal} {Phys.Rev.}\ }\textbf {\bibinfo {volume} {D79}},\ \bibinfo {pages}
  {034502} (\bibinfo {year} {2009})}\BibitemShut {NoStop}%
\bibitem [{\citenamefont {Durr}\ \emph {et~al.}(2008)\citenamefont {Durr},
  \citenamefont {Fodor}, \citenamefont {Frison}, \citenamefont {Hoelbling},
  \citenamefont {Hoffmann} \emph {et~al.}}]{Durr:2008zz}%
  \BibitemOpen
  \bibfield  {author} {\bibinfo {author} {\bibfnamefont {S.}~\bibnamefont
  {Durr}}, \bibinfo {author} {\bibfnamefont {Z.}~\bibnamefont {Fodor}},
  \bibinfo {author} {\bibfnamefont {J.}~\bibnamefont {Frison}}, \bibinfo
  {author} {\bibfnamefont {C.}~\bibnamefont {Hoelbling}}, \bibinfo {author}
  {\bibfnamefont {R.}~\bibnamefont {Hoffmann}},  \emph {et~al.},\ }\href@noop
  {} {\bibfield  {journal} {\bibinfo  {journal} {Science}\ }\textbf {\bibinfo
  {volume} {322}},\ \bibinfo {pages} {1224} (\bibinfo {year}
  {2008})}\BibitemShut {NoStop}%
\bibitem [{\citenamefont {Alexandrou}\ \emph {et~al.}(2009)\citenamefont
  {Alexandrou} \emph {et~al.}}]{Alexandrou:2009qu}%
  \BibitemOpen
  \bibfield  {author} {\bibinfo {author} {\bibfnamefont {C.}~\bibnamefont
  {Alexandrou}} \emph {et~al.},\ }\href@noop {} {\bibfield  {journal} {\bibinfo
   {journal} {Phys.Rev.}\ }\textbf {\bibinfo {volume} {D80}},\ \bibinfo {pages}
  {114503} (\bibinfo {year} {2009})}\BibitemShut {NoStop}%
\bibitem [{\citenamefont {Durr}\ \emph {et~al.}(2012)\citenamefont {Durr},
  \citenamefont {Fodor}, \citenamefont {Hemmert}, \citenamefont {Hoelbling},
  \citenamefont {Frison} \emph {et~al.}}]{Durr:2011mp}%
  \BibitemOpen
  \bibfield  {author} {\bibinfo {author} {\bibfnamefont {S.}~\bibnamefont
  {Durr}}, \bibinfo {author} {\bibfnamefont {Z.}~\bibnamefont {Fodor}},
  \bibinfo {author} {\bibfnamefont {T.}~\bibnamefont {Hemmert}}, \bibinfo
  {author} {\bibfnamefont {C.}~\bibnamefont {Hoelbling}}, \bibinfo {author}
  {\bibfnamefont {J.}~\bibnamefont {Frison}},  \emph {et~al.},\ }\href@noop {}
  {\bibfield  {journal} {\bibinfo  {journal} {Phys.Rev.}\ }\textbf {\bibinfo
  {volume} {D85}},\ \bibinfo {pages} {014509} (\bibinfo {year}
  {2012})}\BibitemShut {NoStop}%
\bibitem [{\citenamefont {Walker-Loud}(2012)}]{WalkerLoud:2011ab}%
  \BibitemOpen
  \bibfield  {author} {\bibinfo {author} {\bibfnamefont {A.}~\bibnamefont
  {Walker-Loud}},\ }\href@noop {} {\bibfield  {journal} {\bibinfo  {journal}
  {Phys.Rev.}\ }\textbf {\bibinfo {volume} {D86}},\ \bibinfo {pages} {074509}
  (\bibinfo {year} {2012})}\BibitemShut {NoStop}%
\bibitem [{\citenamefont {Dai}\ \emph {et~al.}(1996)\citenamefont {Dai},
  \citenamefont {Dashen}, \citenamefont {Jenkins},\ and\ \citenamefont
  {Manohar}}]{Dai:1995zg}%
  \BibitemOpen
  \bibfield  {author} {\bibinfo {author} {\bibfnamefont {J.}~\bibnamefont
  {Dai}}, \bibinfo {author} {\bibfnamefont {R.~F.}\ \bibnamefont {Dashen}},
  \bibinfo {author} {\bibfnamefont {E.~E.}\ \bibnamefont {Jenkins}}, \ and\
  \bibinfo {author} {\bibfnamefont {A.~V.}\ \bibnamefont {Manohar}},\
  }\href@noop {} {\bibfield  {journal} {\bibinfo  {journal} {Phys.Rev.}\
  }\textbf {\bibinfo {volume} {D53}},\ \bibinfo {pages} {273} (\bibinfo {year}
  {1996})}\BibitemShut {NoStop}%
\bibitem [{\citenamefont {Flores-Mendieta}\ \emph {et~al.}(1998)\citenamefont
  {Flores-Mendieta}, \citenamefont {Jenkins},\ and\ \citenamefont
  {Manohar}}]{FloresMendieta:1998ii}%
  \BibitemOpen
  \bibfield  {author} {\bibinfo {author} {\bibfnamefont {R.}~\bibnamefont
  {Flores-Mendieta}}, \bibinfo {author} {\bibfnamefont {E.~E.}\ \bibnamefont
  {Jenkins}}, \ and\ \bibinfo {author} {\bibfnamefont {A.~V.}\ \bibnamefont
  {Manohar}},\ }\href@noop {} {\bibfield  {journal} {\bibinfo  {journal} {Phys.
  Rev. D}\ }\textbf {\bibinfo {volume} {58}},\ \bibinfo {pages} {094028}
  (\bibinfo {year} {1998})}\BibitemShut {NoStop}%
\bibitem [{\citenamefont {Gasser}\ and\ \citenamefont
  {Leutwyler}(1984)}]{Gasser:1983yg}%
  \BibitemOpen
  \bibfield  {author} {\bibinfo {author} {\bibfnamefont {J.}~\bibnamefont
  {Gasser}}\ and\ \bibinfo {author} {\bibfnamefont {H.}~\bibnamefont
  {Leutwyler}},\ }\href@noop {} {\bibfield  {journal} {\bibinfo  {journal}
  {Annals Phys.}\ }\textbf {\bibinfo {volume} {158}},\ \bibinfo {pages} {142}
  (\bibinfo {year} {1984})}\BibitemShut {NoStop}%
\bibitem [{\citenamefont {Ecker}\ \emph {et~al.}(1989)\citenamefont {Ecker},
  \citenamefont {Gasser}, \citenamefont {Leutwyler}, \citenamefont {Pich},\
  and\ \citenamefont {de~Rafael}}]{Ecker:1989yg}%
  \BibitemOpen
  \bibfield  {author} {\bibinfo {author} {\bibfnamefont {G.}~\bibnamefont
  {Ecker}}, \bibinfo {author} {\bibfnamefont {J.}~\bibnamefont {Gasser}},
  \bibinfo {author} {\bibfnamefont {H.}~\bibnamefont {Leutwyler}}, \bibinfo
  {author} {\bibfnamefont {A.}~\bibnamefont {Pich}}, \ and\ \bibinfo {author}
  {\bibfnamefont {E.}~\bibnamefont {de~Rafael}},\ }\href@noop {} {\bibfield
  {journal} {\bibinfo  {journal} {Phys.Lett.}\ }\textbf {\bibinfo {volume}
  {B223}},\ \bibinfo {pages} {425} (\bibinfo {year} {1989})}\BibitemShut
  {NoStop}%
\bibitem [{\citenamefont {Dashen}\ \emph {et~al.}(1994)\citenamefont {Dashen},
  \citenamefont {Jenkins},\ and\ \citenamefont {Manohar}}]{Dashen:1993jt}%
  \BibitemOpen
  \bibfield  {author} {\bibinfo {author} {\bibfnamefont {R.~F.}\ \bibnamefont
  {Dashen}}, \bibinfo {author} {\bibfnamefont {E.~E.}\ \bibnamefont {Jenkins}},
  \ and\ \bibinfo {author} {\bibfnamefont {A.~V.}\ \bibnamefont {Manohar}},\
  }\href@noop {} {\bibfield  {journal} {\bibinfo  {journal} {Phys.Rev.}\
  }\textbf {\bibinfo {volume} {D49}},\ \bibinfo {pages} {4713} (\bibinfo {year}
  {1994})}\BibitemShut {NoStop}%
\bibitem [{\citenamefont {Okun}(2014)}]{Okun_2014}%
  \BibitemOpen
  \bibfield  {author} {\bibinfo {author} {\bibfnamefont {L.~B.}\ \bibnamefont
  {Okun}},\ }\href@noop {} {\emph {\bibinfo {title} {Leptons and Quarks}}}\
  (\bibinfo  {publisher} {{WORLD} {SCIENTIFIC}},\ \bibinfo {year}
  {2014})\BibitemShut {NoStop}%
\bibitem [{\citenamefont {Bernard}\ \emph {et~al.}(2003)\citenamefont
  {Bernard}, \citenamefont {Hemmert},\ and\ \citenamefont
  {Meissner}}]{Bernard:2003xf}%
  \BibitemOpen
  \bibfield  {author} {\bibinfo {author} {\bibfnamefont {V.}~\bibnamefont
  {Bernard}}, \bibinfo {author} {\bibfnamefont {T.~R.}\ \bibnamefont
  {Hemmert}}, \ and\ \bibinfo {author} {\bibfnamefont {U.-G.}\ \bibnamefont
  {Meissner}},\ }\href@noop {} {\bibfield  {journal} {\bibinfo  {journal}
  {Phys.Lett.}\ }\textbf {\bibinfo {volume} {B565}},\ \bibinfo {pages} {137}
  (\bibinfo {year} {2003})}\BibitemShut {NoStop}%
\bibitem [{\citenamefont {Frink}\ and\ \citenamefont
  {Meissner}(2006)}]{Frink:2006hx}%
  \BibitemOpen
  \bibfield  {author} {\bibinfo {author} {\bibfnamefont {M.}~\bibnamefont
  {Frink}}\ and\ \bibinfo {author} {\bibfnamefont {U.-G.}\ \bibnamefont
  {Meissner}},\ }\href@noop {} {\bibfield  {journal} {\bibinfo  {journal} {Eur.
  Phys. J.}\ }\textbf {\bibinfo {volume} {A29}},\ \bibinfo {pages} {255}
  (\bibinfo {year} {2006})}\BibitemShut {NoStop}%
\bibitem [{\citenamefont {Holmberg}\ and\ \citenamefont
  {Leupold}(2018)}]{Holmberg:2018dtv}%
  \BibitemOpen
  \bibfield  {author} {\bibinfo {author} {\bibfnamefont {M.}~\bibnamefont
  {Holmberg}}\ and\ \bibinfo {author} {\bibfnamefont {S.}~\bibnamefont
  {Leupold}},\ }\href@noop {} {\bibfield  {journal} {\bibinfo  {journal} {Eur.
  Phys. J.}\ }\textbf {\bibinfo {volume} {A54}},\ \bibinfo {pages} {103}
  (\bibinfo {year} {2018})}\BibitemShut {NoStop}%
\bibitem [{\citenamefont {Goity}\ \emph {et~al.}(1999)\citenamefont {Goity},
  \citenamefont {Lewis}, \citenamefont {Schvellinger},\ and\ \citenamefont
  {Zhang}}]{Goity:1999by}%
  \BibitemOpen
  \bibfield  {author} {\bibinfo {author} {\bibfnamefont {J.~L.}\ \bibnamefont
  {Goity}}, \bibinfo {author} {\bibfnamefont {R.}~\bibnamefont {Lewis}},
  \bibinfo {author} {\bibfnamefont {M.}~\bibnamefont {Schvellinger}}, \ and\
  \bibinfo {author} {\bibfnamefont {L.-Z.}\ \bibnamefont {Zhang}},\ }\href@noop
  {} {\bibfield  {journal} {\bibinfo  {journal} {Phys. Lett. B}\ }\textbf
  {\bibinfo {volume} {454}},\ \bibinfo {pages} {115} (\bibinfo {year}
  {1999})}\BibitemShut {NoStop}%
\bibitem [{\citenamefont {Gell-Mann}\ \emph {et~al.}(1968)\citenamefont
  {Gell-Mann}, \citenamefont {Oakes},\ and\ \citenamefont
  {Renner}}]{GellMann:1968rz}%
  \BibitemOpen
  \bibfield  {author} {\bibinfo {author} {\bibfnamefont {M.}~\bibnamefont
  {Gell-Mann}}, \bibinfo {author} {\bibfnamefont {R.}~\bibnamefont {Oakes}}, \
  and\ \bibinfo {author} {\bibfnamefont {B.}~\bibnamefont {Renner}},\
  }\href@noop {} {\bibfield  {journal} {\bibinfo  {journal} {Phys.Rev.}\
  }\textbf {\bibinfo {volume} {175}},\ \bibinfo {pages} {2195} (\bibinfo {year}
  {1968})}\BibitemShut {NoStop}%
\bibitem [{\citenamefont {Dashen}\ and\ \citenamefont
  {Weinstein}(1969)}]{Dashen:1970vh}%
  \BibitemOpen
  \bibfield  {author} {\bibinfo {author} {\bibfnamefont {R.~F.}\ \bibnamefont
  {Dashen}}\ and\ \bibinfo {author} {\bibfnamefont {M.}~\bibnamefont
  {Weinstein}},\ }\href@noop {} {\bibfield  {journal} {\bibinfo  {journal}
  {Phys. Rev.}\ }\textbf {\bibinfo {volume} {188}},\ \bibinfo {pages} {2330}
  (\bibinfo {year} {1969})}\BibitemShut {NoStop}%
\bibitem [{\citenamefont {Heo}\ and\ \citenamefont {Lutz}(2018)}]{Heo:2018mur}%
  \BibitemOpen
  \bibfield  {author} {\bibinfo {author} {\bibfnamefont {Y.}~\bibnamefont
  {Heo}}\ and\ \bibinfo {author} {\bibfnamefont {M.~F.~M.}\ \bibnamefont
  {Lutz}},\ }\href@noop {} {\bibfield  {journal} {\bibinfo  {journal} {Phys.
  Rev.}\ }\textbf {\bibinfo {volume} {D97}},\ \bibinfo {pages} {094004}
  (\bibinfo {year} {2018})}\BibitemShut {NoStop}%
\bibitem [{\citenamefont {Lutz}\ \emph {et~al.}(2020)\citenamefont {Lutz},
  \citenamefont {Sauerwein},\ and\ \citenamefont {Timmermans}}]{Lutz:2020dfi}%
  \BibitemOpen
  \bibfield  {author} {\bibinfo {author} {\bibfnamefont {M.~F.}\ \bibnamefont
  {Lutz}}, \bibinfo {author} {\bibfnamefont {U.}~\bibnamefont {Sauerwein}}, \
  and\ \bibinfo {author} {\bibfnamefont {R.~G.}\ \bibnamefont {Timmermans}},\
  }\href@noop {} {\bibfield  {journal} {\bibinfo  {journal} {Eur. Phys. J. C}\
  }\textbf {\bibinfo {volume} {80}},\ \bibinfo {pages} {844} (\bibinfo {year}
  {2020})}\BibitemShut {NoStop}%
\bibitem [{\citenamefont {Zyla}\ \emph {et~al.}(2020)\citenamefont {Zyla} \emph
  {et~al.}}]{Zyla:2020zbs}%
  \BibitemOpen
  \bibfield  {author} {\bibinfo {author} {\bibfnamefont {P.}~\bibnamefont
  {Zyla}} \emph {et~al.},\ }\href@noop {} {\bibfield  {journal} {\bibinfo
  {journal} {PTEP}\ }\textbf {\bibinfo {volume} {2020}},\ \bibinfo {pages}
  {083C01} (\bibinfo {year} {2020})}\BibitemShut {NoStop}%
\bibitem [{\citenamefont {Guo}\ \emph {et~al.}(2020)\citenamefont {Guo},
  \citenamefont {Heo},\ and\ \citenamefont {Lutz}}]{Guo:2019nyp}%
  \BibitemOpen
  \bibfield  {author} {\bibinfo {author} {\bibfnamefont {X.-Y.}\ \bibnamefont
  {Guo}}, \bibinfo {author} {\bibfnamefont {Y.}~\bibnamefont {Heo}}, \ and\
  \bibinfo {author} {\bibfnamefont {M.~F.~M.}\ \bibnamefont {Lutz}},\
  }\href@noop {} {\bibfield  {journal} {\bibinfo  {journal} {Eur. Phys. J. C}\
  }\textbf {\bibinfo {volume} {80}},\ \bibinfo {pages} {260} (\bibinfo {year}
  {2020})}\BibitemShut {NoStop}%
\end{thebibliography}%
\end{document}